\documentclass[lettersize,journal]{IEEEtran}
\IEEEoverridecommandlockouts

\usepackage{url}
\usepackage{times}
\usepackage{amsmath}
\usepackage{epsfig}
\usepackage{multirow}
\usepackage{longtable}
\usepackage{amsfonts}
\usepackage{amssymb}%
\usepackage{color}
\usepackage{bm}
\usepackage{epstopdf}
\usepackage{subfigure}
\usepackage{cite}
\usepackage{graphicx}
\usepackage{amsmath}
\usepackage{times}
\usepackage{latexsym}
\usepackage[center]{caption2}
\usepackage{stfloats}
\usepackage{cases}
\usepackage{array}
\usepackage{setspace}
\usepackage{fancyhdr}
\usepackage{balance}
\usepackage{algorithm} 
\usepackage{algpseudocode} 
\usepackage{multicol}
\usepackage[table]{xcolor}
\usepackage{makecell}

        {\par\noindent{\bf Proof.}}%
        {{\hfill{\large$\Box$}}\smallskip}

\begin{document}


\title{When Learning Meets Dynamics: Distributed User Connectivity Maximization in UAV-Based Communication Networks

}


\author{\IEEEauthorblockN{
Bowei Li\IEEEauthorrefmark{1}, 
Saugat Tripathi\IEEEauthorrefmark{2},
Salman Hosain\IEEEauthorrefmark{3},
Ran~Zhang\IEEEauthorrefmark{3},~\IEEEmembership{Senior Member,~IEEE,}
Jiang (Linda) Xie\IEEEauthorrefmark{3},~\IEEEmembership{Fellow,~IEEE,} and
Miao~Wang\IEEEauthorrefmark{3},~\IEEEmembership{Senior Member,~IEEE}
}\\
~\\
\IEEEauthorblockA{\IEEEauthorrefmark{1}Department of Electrical and Computer Engineering, Carnegie Mellon University, Pittsburgh, PA, USA}\\
\IEEEauthorblockA{\IEEEauthorrefmark{2}Department of Sensor Engineering, PassiveLogic Inc., Salt Lake City, UT, USA,}\\
\IEEEauthorblockA{\IEEEauthorrefmark{3}
Department of Electrical and Computer Engineering, 
University of North Carolina at Charlotte, NC, USA}\\
\IEEEauthorblockA{Email: 
\IEEEauthorrefmark{1}boweili@andrew.cmu.edu, \IEEEauthorrefmark{2}saugat@passivelogic.com, \IEEEauthorrefmark{3}\{$ahosain,rzhang8,linda.xie,mwang25$\}@charlotte.edu}
\thanks{This work was partly presented at 2023 IEEE GLOBECOM \cite{previouswork}.} }

\maketitle



\thispagestyle{fancy}
\fancyhead[C]{\small{\selectfont This work has been submitted to the IEEE for possible publiction. Copyright may be transferred without notice, after which this version may no longer be accessible.}}

\renewcommand{\headrule}{}
\pagestyle{plain}

\begin{abstract}
Distributed management over Unmanned Aerial Vehicle (UAV) based communication networks (UCNs) has attracted increasing research attention. In this work, we study a distributed user connectivity maximization problem in a UCN. The work features a horizontal study over different levels of information exchange during the distributed iteration and a consideration of dynamics in UAV set and user distribution, which are not well addressed in the existing works. Specifically, the studied problem is first formulated into a time-coupled mixed-integer non-convex optimization problem. A heuristic two-stage UAV-user association policy is proposed to faster determine the user connectivity. To tackle the NP-hard problem in scalable manner, the distributed user connectivity maximization algorithm 1 (DUCM-1) is proposed under the multi-agent deep Q learning (MA-DQL) framework. DUCM-1 emphasizes on designing different information exchange levels and evaluating how they impact the learning convergence with stationary and dynamic user distribution. To comply with the UAV dynamics, DUCM-2 algorithm is developed which is devoted to autonomously handling arbitrary quit's and join-in's of UAVs in a considered time horizon. Extensive simulations are conducted \textit{i)} to conclude that exchanging state information with a deliberated task-specific reward function design yields the best convergence performance, and \textit{ii)} to show the efficacy and robustness of DUCM-2 against the dynamics. 

\end{abstract}
\section{Introduction}\label{sec.Intro}

Unmanned Aerial Vehicle (UAV) based communication networks (UCN) have been envisioned as one of the indispensable components in future mobile communications. Taking advantage of the swift 3D mobility, potentially better air-to-ground channel, and low deployment and operational cost, UCNs can provide highly on-demand communication services to ground and aerial users as a preeminent enhancement or complement to existing telecommunication infrastructure\cite{zeng2016wireless}. 

UCNs have been studied from various aspects\cite{c}.
A diversity of approaches have been applied in either a centralized or distributed manner. Particularly, a centralized decider needs to collect the complete network information and makes decisions for all the UAVs. While possibly gaining better global performance, centralized approaches require strong computing power from a single decider, and unacceptable communication overhead and latency may be incurred during information collection and decision dissemination\cite{ei2022energy,zhang2019cellular,li2022network,ConsKha,luong2021deep,liu2018energy}. 
On the contrast, decentralized approaches distribute the computing and communication load to individual UAVs such that each UAV can iteratively incubate and ameliorate its own policy based on its local observations and light-weight information exchange with other UAVs\cite{chai2021multi,trotta2018joint,liu2019task,hu2021distributed,cui2019multi,park2022cooperative,pham2018cooperative}. With increasing UCN size and the desire for lower communication overhead and latency, distributed approaches are becoming a better choice to achieve scalability.

In a distributed design, UAVs coordinate for a shared goal via iteratively exchanging information and updating their own policies for a final convergence. The design of information to exchange is the key to a successful convergence. An intuitional horizontal comparison between different levels of information exchange will provide valuable insights on how to design a more efficient distributed algorithm and better tune the tradeoff between convergence performance and implementation complexity. Nevertheless, few existing works have dabbled in the impact of different exchange levels on the convergence performance in a UCN context. In addition, in a DCN, dynamics are common for the involved UAVs and users. The user distribution may vary both temporally and spatially in the considered time horizon. The UAV crew may also dynamically change during the mission as \textit{i)} existing UAVs may run out of battery and thus need to quit for charging, or \textit{ii)} supplementary UAVs may be dispatched to join the existing crew for performance enhancement\cite{zhang2020srec}. However, not many works have considered the adaptability of the proposed approaches to the above dynamics, especially to the dynamic UAV set. Some prior works such as \cite{kimura2020distributed,zhang2021learning,tang2020deep,wang2023optimal,liu2019trajectory} did consider such dynamics to some extent, but they either adopted centralized approaches or considered a fixed UAV set. Adaptive, distributed DCN management strategies are still much desired to handle the dynamic change in UAV set and user distributions in a scalable manner. 

To this end, we study a distributed user connectivity maximization problem, where a UCN is deployed over a target region to provide connectivity services to the ground users. The UAV crew are set to change dynamically with arbitrary (not pre-planned) quit's and join-in's in a time horizon. The user distribution may also change temporally and spatially across the region. The objectives are to \textit{i)} develop a distributed approach to obtain a UAV trajectory control strategy that can maximize the number of connected users in a time horizon\footnote{Note that a user is in the coverage of a UAV does not necessarily mean it can connect to the UAV. The studied problem is not simply a maximum dot covering problem.} while taking care of the above dynamics; \textit{ii)} evaluate the impact of different levels of inter-UAV information exchange on the convergence performance of the designed approach. Multi-agent reinforcement learning (MARL) is applied as the distributed framework\cite{zhang2021multi} among a variety of distributed algorithms\cite{chai2021multi,trotta2018joint,liu2019task,yang2021privacy,zhang2020federated,hu2021distributed,cui2019multi,park2022cooperative,pham2018cooperative}.
Compared to others, MARL has strong capability of making time-sequential decisions in a dynamic environment with limited domain knowledge\cite{hou2023uav}. The distributed learning agents iteratively update their own strategies via deliberate trial-and-error and reward feedback instead of following predefined rules, thus being free of the environment model. This learning feature also makes the achieved strategy more robust against the system dynamics. A well-trained model can not only provide the best action for any situation it has ever seen but also adapt faster to unseen situations via transfer learning rather than compute from scratch as many conventional approaches do. Specifically, our contributions are summarized as follows.
\begin{itemize}
    \item The user connectivity maximization problem is formulated into a time-coupled mixed-integer non-convex optimization problem. In the optimization, a heuristic two-stage user admission and association policy is proposed to fast determine the user connectivity to the UAVs.
    \item To tackle the NP-hard problem in a scalable manner, the distributed user connectivity maximization algorithm 1 (DUCM-1) is first designed under the multi-agent deep Q learning (MA-DQL) framework. The approach of centralized training and distributed execution is adopted\cite{chen2019new}. As the first step, DUCM-1 considers fixed serving UAV crew but emphasizes on evaluating how different information exchange levels and reward function designs impact the learning convergence. Four different designs are proposed with discussions on their potentials. The best design will be applied to further help accommodate the UAV dynamics.  
    \item To comply with the UAV dynamics, DUCM-2 algorithm is developed exploiting the best design from DUCM-1. DUCM-2 is capable of autonomously handling arbitrary quit's and join-in's of UAVs in a considered time horizon. The algorithm features a heterogeneous episode design for generalization, parallel complementary training to improve time efficiency, and can obtain a distributed and adaptive policy with just one single training.
    \item Extensive simulations are conducted to \textit{i)} evaluate the convergence performance under different designs of information exchange and reward function in cases of stationary and dynamic user distributions, \textit{ii)} demonstrate the efficacy and robustness of DUCM-2 against the dynamic change in UAV set and their starting positions, and \textit{iii)} provide example UAV trajectories for case study.

\end{itemize}

The remainder of the paper is organized as follows. Section II presents the related works. Section III describes the system model. Section IV provides the problem formulation. Section V and VI detail the design of the proposed DUCM algorithms, with emphasis on different information exchange levels and how to handle arbitrary UAV quit's and join-in's, respectively. Section VII shows the numerical results. Section VI concludes the paper.

\section{Related Works}\label{sec.RelatedWorks}
UCNs have been studied in the existing literature from various aspects in either centralized or distributed manner. The works can be further divided into conventional approaches based on optimization\cite{c}, game theory\cite{mkiramweni2019survey} or statistical analysis\cite{galkin2019stochastic}, and learning based approaches based on advanced machine learning techniques such as federated deep learning (DL)\cite{qu2021decentralized}, reinforcement learning (RL)\cite{bai2023towards} and generative AI (GAI)\cite{liu2024generative}.

The conventional approaches usually exploit known environmental models or domain knowledge to achieve the objectives\cite{ei2022energy,zhang2019cellular,li2022network,chai2021multi,trotta2018joint,liu2019task}. \textit{For centralized approaches}, the originally formulated problems are usually mixed-integer non-convex optimizations. They are generally intractable, but can be resolved using relaxation or decomposition approaches. For instance, Ei \textit{et al.}\cite{ei2022energy} utilized a relaxed block successive upper-bound minimization method; Zhang \textit{et al.}\cite{zhang2019cellular} decoupled the original problem into subproblems which were solved by an iterative algorithm; Li \textit{et al.}\cite{li2022network} adopted an alternative optimization and Dinkelbach's method. \textit{For distributed approaches}, Chai \textit{et al.}\cite{chai2021multi} carried forward a novel joint trajectory and communication scheduling scheme for UAV-enabled wireless caching networks. The problem was formulated as an infinite horizon ergodic stochastic differential game. Mean-field equilibrium analysis was exploited to obtain a decentralized solution. Trotta \textit{et al.}\cite{trotta2018joint} discussed joint UAV coverage maximization and ground charging scheduling for a UAV network. Normal-form games were adopted to make the UAV charging scheduling distributed. A bio-inspired scheme was used for distributed UAV positioning. Liu \textit{et al.}\cite{liu2019task} considered a distributed UAV relay assignment problem in a UCN where congestion game model was utilized to self-organize coordination among source UAVs and relay UAVs.

The learning based approaches do not require preliminary knowledge of the environmental models since the agents can learn that during training via a huge data set or constant interaction with the environment. These merits make them applicable to more complex and dynamic scenarios, yet more prune to the hyperparameters of the learning algorithms\cite{liu2020artificial}. \textit{Centralized learning approaches} applied to UCNs are mostly RL-based due to its strong capability of decision making in model-free and dynamic environment. For instance, Khairy \textit{et al.}\cite{ConsKha}, Luong \textit{et al.}\cite{luong2021deep} and Liu \textit{et al.} \cite{liu2018energy} studied joint altitude control and channel access management, joint UAV positioning and radio resource allocation, and energy-efficient coverage maximization problems in UAV networks, using Proximal Policy Optimization (PPO), deep Q learning (DQL), and deep deterministic policy gradient (DDPG) algorithms, respectively. More recently, GAI has received increasing attention due to its powerful learning and generalization capabilities. A few works\cite{liu2024generative,sun2024generative,sharif2024resource} have been discussing the potential applications of GAI in UAV networks in terms of channel estimation, interactive strategy optimization, routing, topology design, etc. 

As to \textit{distributed learning approaches}, federated deep learning has been adopted extensively to learn a common model in a distributed manner, allowing scalability and privacy-preserving\cite{yang2021privacy,zhang2020federated}. MARL has also been extensively applied, where UAVs as agents can learn their individual policies based on local observations and iterative information exchange. For instance, Hu \textit{et al.} \cite{hu2021distributed} studied the trajectory design problem to cooperatively serve ground users in uplink. UAVs share their individual states and each UAV makes decisions based on the action history of the other UAVs. Cui \textit{et al.} \cite{cui2019multi} adopted multi-agent Q learning to tackle a joint user selection and radio resource allocation problem where UAVs implicitly interact with each other via user selection. Park \textit{et al.} \cite{park2022cooperative} proposed a multi-UAV positioning algorithm under the multi-agent actor-critic framework. UAVs share individual rewards to train their own neural networks. Pham \textit{et al.} \cite{pham2018cooperative} developed a multi-agent correlated Q learning algorithm to guide UAVs to cover an area with minimum overlapping. UAVs share individual state information to calculate task-specific team rewards.

The existing MARL based works have not considered how different levels of information sharing will impact the learning convergence. In addition, not many works have taken into account the dynamics of the user distribution and UAV set during the considered time horizon. Some prior works \cite{kimura2020distributed,zhang2021learning,tang2020deep,wang2023optimal,liu2019trajectory} considered such dynamics to some extent, but they either adopted centralized approaches\cite{zhang2021learning,tang2020deep,wang2023optimal,liu2019trajectory} or considered a fixed UAV set\cite{kimura2020distributed,tang2020deep,liu2019trajectory}. Our prior work\cite{previouswork} discussed that info-sharing level would affect the learning convergence, but still assumed a fixed UAV set.

\section{System Model}\label{sec.SystemModel}
\begin{figure}[!ht]
	\centering
	\includegraphics[width=3.0in]{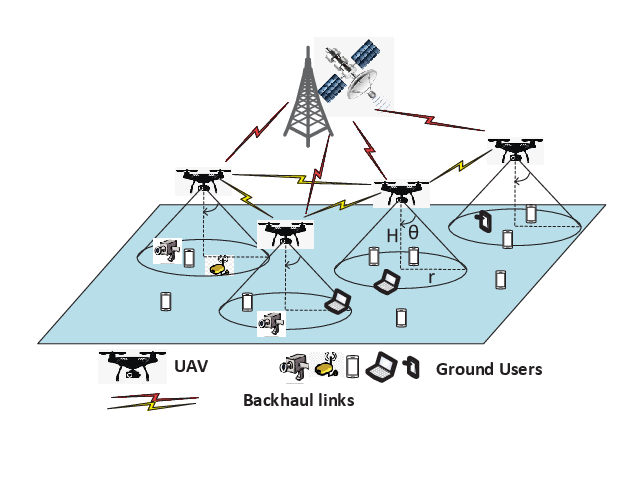}
	\vspace{-1cm}
    \caption{Network Model.} \label{fig.SystemModel}
\end{figure}

\subsection{Network Model}\label{subsec.NetMdl}
As shown in Fig. \ref{fig.SystemModel}, we consider a set of UAVs, denoted as $\mathcal{I}_t$, flying over an $L\times L$ target region at an altitude $H$ to provide communication services to the ground users. The considered time horizon $T$ is slotted into $t$-indexed time steps. The UAV set $\mathcal{I}_t$ may change at any $t$ during the considered time horizon. Each UAV is equipped with directional antennas so that most of its transmission energy is concentrated on an aperture angle of $\theta$ underneath the UAV. As a result, the ground coverage of each UAV is a disk with radius $r=H\cdot \tan(\theta/2)$. There are $U$ users distributed in the region. A percentage of $p$ users are distributed in hot spots while the remaining are uniformly distributed throughout the region. Denote the set of users as $\mathcal{U}_t$, which might change at any $t$. The positions of UAVs and users at time $t$ are denoted by 2D-cartesian coordinates $(x_{i,t}, y_{i,t})$ and $(x_{u,t}, y_{u,t})$, respectively, which are bounded by
\begin{equation}\label{eq.bndry}
0\leq x_{i,t}, y_{i,t}, x_{u,t}, y_{u,t} \leq L,~\forall i\in\mathcal{I}_t,~t\in\{1,2,\cdots,T\}.
\end{equation} 

We consider that UAVs fly without human intervention or a centralized control unit. Each UAV makes individual time-sequential decisions on its own movements along time steps $t$ autonomously yet coordinately. The decisions are based on its local observations (or individual states) and/or the exchanged information with other UAVs. 


\subsection{Spectrum Access}\label{subsec.sa}
All UAVs share the same spectrum. Users follow the way of Orthogonal Frequency Division Multiple Access (OFDMA) to access the spectrum of UAVs: The available bandwidth of individual UAVs is divided into orthogonal resource blocks (RBs) such that users occupying different RBs of the same UAV have no mutual interference. When connecting to UAV $i$, user $u$ is assigned with a certain number of RBs $N_{i,u}$ based on its throughput requirement and channel conditions. We consider that each user has a minimum throughput requirement $r_{min}$. Therefore, $N_{i,u}$ can be determined as the smallest integer that satisfies the following condition:
\begin{equation}\label{eq.sa}
r_{i,u} = \sum_{n=1}^{N_{i,u}} BW_{rb} \log_2(1+SINR_{iu,n}) \geq r_{min},
\end{equation} 
where $r_{i,u}$ represents the actual throughput of the user, and $SINR_{iu,n}$ denotes the signal-to-interference-and-noise ratio (SINR) at user $u$ from the connected UAV $i$ on the $n$-th assigned RB. The interference comes from other UAVs than $i$ which cover user $u$ and transmit on the same RBs as assigned to user $u$. When the UAVs are sufficiently dispersed and there is little overlapping between their coverage disks, the $SINR_{iu,n}$ will be simplified into $SNR_{iu}$, where only noise matters. Particularly, $SINR_{iu,n}$ is given as
\begin{equation}\label{eq.snir}
\begin{split}
&SINR_{iu,n} = \frac{P_t G_{iu}}{n_0 + \sum\limits_{\substack{{j \in {\mathcal{I}}_{u,n}} \\ {j \neq i}}} P_t G_{ju}},\\ 
&\text {where }G_{iu}=10^{-PL_{iu}/20}.
\end{split}
\end{equation} 
Here, $P_t$ denotes the transmit power spectrum density (PSD) of UAVs, $n_0$ denotes the PSD of noise at user $u$, and $G_{iu}$ denotes the channel gain at user $u$ from UAV $i$. The set $\mathcal{I}_{u,n}$ denotes the set of UAVs that cover user $u$ and transmit on $n$th RB. The symbol $PL_{iu}$ denotes the air-to-ground path loss which is composed of free-space path loss and excessive path loss determined by the propagation environment\cite{ModelingAl}: 
\begin{equation}\label{eq.pl}
\begin{split}
PL_{iu} = 20\log _{10}{\left({\frac {4\pi f_{c}d_{iu}}{c}}\right)}+\eta \ \ \  \text{(dB),}
\end{split}
\end{equation} 
where $f_c$, $d_{iu}$, $c$, and $\eta$ represents the center carrier frequency, the distance between UAV $i$ and user $u$, the light speed, and excessive path loss, respectively. Particularly, $\eta$ takes different values for LoS/non-LoS channels and various urban environments.

\subsection{User Admission and Association}\label{subsec.userassl}
A user $u$ is considered connected if it is admitted and assigned with RBs by a UAV $i$. Denote the connectivity at time \(t\) as \(X_{u, i}(t)\), which is a binary variable taking value $1$ if user \(u\) is connected to UAV \(i\) and $0$ otherwise. A user can connect to at most one UAV at any time $t$, i.e.,
\begin{equation}\label{eq.connectivity}
\sum_{i \in \mathcal{I}_t}{X_{u, i}(t)} \leq 1 , \; \forall{\: u \in \mathcal{U}_t} \text{ and }t=\{1,2,\cdots,T\}.
\end{equation} 

Meanwhile, the total number of assigned RBs of a UAV should not exceed a cap $N_{rb}$ bounded by the bandwidth, i.e., 
 \begin{equation}\label{eq.totrb}
\sum_{u \in {U_{i,t}}}{N_{i,u}} \leq N_{rb} ,\;\forall i\in \mathcal{I}_t,
\end{equation} 
where \(U_{i,t} \subset \mathcal{U}_t\) denotes the set of users connected to UAV \(i\) at time \(t\).

A two-step user association policy is adopted. In the first step, each user, if covered by any UAVs, first sends a connection request to the UAV that provides the best average air-to-ground channel gain. Each UAV then sorts all the received connection requests and admits the users in a descending order of channel gains subject to its RB availability. The RBs of a UAV are assigned in the natural number indexing order. In the second step, each unadmitted user finds its next best UAV and sends a connection request. The request sending is repeated until the user is admitted by a UAV or runs out of UAVs to send request to. 
Note that such a policy is heuristic and can be sub-optimal when UAVs overlap considerably and RB allocation can be coordinated among UAVs to further increase the overall connectivity. But when the UAVs are sufficiently dispersed, the gap to the optimality is marginal. As the emphasis of the work is on distributed UAV trajectory design to accommodate the dynamic change in the UAV set and user distribution, we argue that adopting a heuristic user association policy will not affect the design efficacy in solving the major concern.

\section{Problem Formulation}\label{sec.PF}
The objective of this work is to obtain a trajectory control policy to optimally guide the UAVs to move from the initial positions to those that maximize the total number of served users. The policy is expected to be robust against the dynamic change in the serving UAV crew and user distribution as well as the initial positions of the UAVs. The decision variables for the optimization will be the movements of all the UAVs at each time step within the considered time horizon, denoted as $\{a_{i,t}\}, \text{where} \;\forall i\in\mathcal{I}_t,\;t=\{1,\cdots,T\}$, and $\mathcal{I}_t$ denotes the serving UAV set at time step $t$. The problem formulation is given as follows.
In the problem $P$, $\mathcal{U}_{i,t}$ denotes the set of users connected to UAV $i$ at time step $t$. The user connectivity $\{X_{u, i}(t)\}$, user sets $\{\mathcal{U}_{i,t}\}$ and assigned bandwidth $\{N_{i,u}\}$ are jointly determined by the UAV positions $(x_{i,t},y_{i,t})$ at time step $t$ and the user association policy outlined in Subsection \ref{subsec.userassl}. Individual UAV positions $\{(x_{i,t},y_{i,t})\}$ are further determined by the accumulated UAV movements $\{a_{i,t}\}$ up to time step $t$. The constraints C1, C2, C3, and C4 are for the user association, multi-connectivity, available UAV bandwidth, and boundary limitation, respectively. 
\begin{alignat}{2}
&\max_{a_{i,t},\;\forall i\in \mathcal{I}_t,\;t\in \{1,\cdots,T\}}\left[\sum_{t=1}^T\sum_{i \in \mathcal{I}_t}{\sum_{u \in \mathcal{U}_{i,t}}{X_{u, i}(t)}}\right] \tag{P}\\
\mbox{s.t.} & \quad X_{u, i}(t) \in \{0, 1\}, \forall{i \in \mathcal{I}_t \text{ and } \forall u \in \mathcal{U}_{i,t}} \tag{C1} \\
&\quad \sum_{i \in \mathcal{I}_t}{X_{u, i}(t)} \leq 1 , \; \forall{\: u \in \mathcal{U}_{i,t}} \tag{C2}\\
&\quad\sum_{u \in {\mathcal{U}_{i,t}}}{N_{i,u}} \leq N_{rb}, \forall{i \in \mathcal{I}_t} \tag{C3}\\
&\quad 0 \leq x_{i,t}, \;y_{i,t} \leq L, \forall{i \in \mathcal{I}_t} \tag{C4}
\end{alignat}

The formulated problem is a mixed-integer nonlinear non-convex problem with nonlinear constraints. The items of different $t$ in the objective function are temporally coupled through the time-sequential UAV positions. These facts make the sequential decision problem NP-hard and intractable. In addition, the achieved policy by solving the optimization problem directly is parameter-sensitive, that is, the optimization needs to be re-executed every time when the network parameters are updated. This will make the direct optimization implementation-unfriendly as the UAV networks are highly dynamic. To this end, we propose to crack the problem using DRL methods which are strongly capable of making time-sequential decisions in a dynamic environment. The multi-agent framework will be exploited to make the learning and policy execution distributed for better scalability and robustness. Note that RL methods can guarantee convergence if designed properly but not the optimality. Therefore, how to design the learning algorithms to promote exploration for the better policy in an efficient way is the key to success.
\section{Design of Distributed User Connectivity Maximization Algorithm 1 (DUCM-1 Algorithm)}\label{sec.Algorithm1}
This section details the design of DUCM-1 algorithm when the serving UAV crew do not change in the considered time horizon. The emphasis is to evaluate how different designs on information exchange level and reward function impact the learning convergence. The best design will be exploited to further help accommodate the dynamic change in the UAV set and user distribution, which is mostly elaborated in the next section. The algorithm is designed under the MA-DQL framework, where each UAV is considered as an agent. They have their own local state and action space and individual rewards. Each UAV trains and executes its own policy, represented by a deep neural network, to determine its next move (i.e., action) based on its local states and shared information from other UAVs.

\subsection{State Space}\label{subsec.statespace}
Denote the local state space for UAV $i$ as $\mathcal{S}_i$. The entire target region is discretized into an $M$-by-$M$ grid with equal spacing. The positions of the UAVs are constrained to the grid intersections. The UAV position is part of the local states since it determines UAV's user connectivity. In addition, when the user distribution changes dynamically, time slot index $t$ is also included as one of the states to help the algorithm better capture the user dynamics. Therefore, the local state of UAV $i$ at time $t$, i.e., $s_{i,t}$, is defined as 
\begin{equation}\label{eq.state}
s_{i,t} = (x_{i,t},\;y_{i,t},\;t),\;\forall i\in\mathcal{I}_t,\; t = \{1,\cdots,T\},
\end{equation}
where state $t$ is optional depending on whether dynamic user distributions are considered. For a standard Q learning (QL) design, the total number of possible local states for one UAV is $M^2T$. Yet for a DQL design, the input dimension to the deep Q Network (DQN) is equal to the dimension of the local state space. Thus, the DQN will have 3 inputs, one for $x_{i,t}$, one for $y_{i,t}$, and one for step $t$.

\subsection{Action Space}\label{subsec.actionspace}
At any grid intersection, one UAV has five possible horizontal movements: hover, left, right, forward and backward. When a UAV moves, it will move by one grid length. Denote the local action space of UAV $i$ as \(\mathcal{A}_i\), which is given as, 
\begin{equation}\label{eqn8}
\mathcal{A}_i = \{0, 1, 2, 3, 4\},
\end{equation} 
where \(0, 1, 2, 3, 4\) represents the respective horizontal movement. Action \(a_{i,t}\) of UAV \(i\) takes a value from $\mathcal{A}_i$ in each time step $t$. They determine the movements of each UAV over the entire time horizon, changing the respective states accordingly.

\subsection{Design of Information Exchange and Reward Function}\label{subsec.infosharing}
In this subsection, 4 different information exchange levels during the training procedure are presented. Their impact on the learning convergence will be investigated via simulations. For different information exchange levels, the reward function is designed accordingly. Generally speaking, the instantaneous reward of a UAV in step $t$ is composed of three major parts: the reward as a function of the total number of connected users, the reward related to the specific shared information (optional), and the individual penalties for going out of bound.

\subsubsection{Implicit Information Exchange (Level 1)}
In this level, UAVs do not explicitly communicate with each other. However, they indirectly interact through user association. The user already associated with one UAV will not connect to other UAVs. This will make some UAVs aware that the current position may not be good enough and going to other places may let it collect more users. We refer to this kind of implicit interaction as implicit information exchange among UAVs. As such, the reward function of UAV $i$ will only contain the number of its connected users and the penalty for going out of bound, i.e., 
\begin{equation}\label{eqn9}
R_{i} (t) = \sum_{u \in \mathcal{U}_t} {X_{u,i}(t)} - f_i, \: \forall \; {i \; \in \; \mathcal{I}_t}
\end{equation}  
where \(f_i\) denotes the out-of-bound penalty for UAV \(i\). 

\subsubsection{Exchange of Individual Reward Information (Level 2)}
In this level, UAVs share their local user connectivity with others. The individual UAV reward is calculated by averaging the local user connectivities (including its own) over all the UAVs with considering its own out-of-bound penalty. Note that UAVs do not share the individual penalties. Thus, the reward function for UAV $i$ at time step $t$ is calculated as
\begin{equation}\label{eqn10}
\begin{split}
R_{i}(t) &= \frac{\sum\limits_{u \in \mathcal{U}_t,\; j \in \mathcal{I}_{t}} X_{u,j}(t)}{|\mathcal{I}_{t}|} - f_{i}, \: \forall \: i \: \in \: \mathcal{I}_t.
\end{split}
\end{equation} 

\subsubsection{Exchange of Problem-Specific Information (Level 3)} 
In this level, instead of sharing the local user connectivity, each UAV shares its updated position. This shared information will be used in calculating the individual rewards of each UAV. With the position information, the individual reward function will penalize a UAV on getting too close to other UAVs. This will promote the UAVs to disperse as appropriate and reduce their overlapping (thus the competition for users). The distance-based penalty of UAV $i$ with respect to UAV $j$ is calculated as 
\begin{equation}\label{eq.penalty}
\begin{split}
&p_{i,j}(t) = max\left(0, \;(1-\frac{d_{i,j}(t)}{2r}) \cdot p_{max}\right), \\
&\text{where } p_{max} = d_p \cdot \frac{|\mathcal{I}_t|}{|\mathcal{U}_t|}, \;\forall{i,j \in \mathcal{I}_t},\text{} i\ne j
\end{split}
\end{equation}
In Eq. \eqref{eq.penalty}, \(r, \;p_{max},\;d_{i,j}(t)\) denotes the ground coverage radius of a UAV, the maximum distance-based penalty, and the distance between UAV $i$ and UAV $j$ at time $t$, respectively. When calculating \(p_{max}\), $d_p$ is a constant weighting factor that adjusts the impact of $p_{i,j}(t)$ on the individual UAV reward $R_i(t)$. The maximum penalty $p_{max}$ is also tuned by the UAV-user ratio as a higher ratio indicates potentially more intensive competition for users. According to Eq. \eqref{eq.penalty}, if one UAV is at least $2r$ distance away from all the other UAVs, it will not get penalized. With $p_{i,j}(t)$, the individual reward is composed of 3 parts, i.e., 
\begin{equation}\label{eqn11}
R_i(t) = \sum_{u \in \mathcal{U}_t} {X_{u,i}(t)} - \sum_{j \in \mathcal{I}_t, j\ne i}p_{i,j}(t) - f_i, \; \forall{i \in \mathcal{I}_t}.
\end{equation}
Compared to Level-2, level-3 agents yielding low local rewards may know better where to move since they can move away from any nearby UAVs to reduce competing for users. This may lead to a faster convergence or a better overall user connectivity at the end. However, it may not provide satisfying results in occasions where users are so densely distributed in a small area that one UAV cannot cover most of the users there.
\begin{algorithm}[h]
\caption{Distributed User Connectivity Maximization Algorithm 1 (DUCM-1)}\label{alg:madqn}
\begin{algorithmic}[1]
\State \textbf{/*Initialization:*/}
\ForAll{$i \in \mathcal{I}_t$}
    \State Randomly initialize the main DQN $Q(\mathcal{S}_i,\mathcal{A}_i|\theta_i$);
    \State Initialize the target DQN $Q'(\mathcal{S}_i,\mathcal{A}_i|\theta'_i)$ with $\theta_i$;
    \State Initialize experience replay buffer $\mathcal{B}_i$;
\EndFor
\State \textbf{/*Training:*/}
\For{episode := 1:$N_{Eps}$}
    \State {Reset all UAVs to initial states $s_{i,0},\forall i\in\mathcal{I}_t$;}
    \For{time step $t$ := $0:T$}
        \ForAll{$i \in \mathcal{I}_t$}
            \State {/*$\epsilon$-greedy action selection*/}
            \State $\beta\gets$ rand(0,1); 
            \If {$\beta\le\epsilon$}
                \State {$a_{i,t} \gets$ rand([0,1,2,3,4]);}
            \Else
                \State {$a_{i,t} \gets \arg\max_{a_{i}} Q(s_{i,t},a_i|\theta_i)$;}
            \EndIf
            \State {Execute $a_{i,t}$ and observe the next state $s_{i,t+1}$;}
            \State {Share information with other UAVs according}
            \State {to Level 1, 2, 3, or 4;}
            \State {Calculate instantaneous reward $R_i(t+1)$ }
            \State {according to Eq. \eqref{eqn9}, \eqref{eqn10}, or \eqref{eqn11};}
            \State {Store the experience $\{s_{i,t},a_{i,t},s_{i,t+1},R_i(t+$ }
            \State {$1)\}$ into Buffer $\mathcal{B}_i$;}
            \State {/*Update of DQNs*/}
            \State Randomly sample a mini-batch from $\mathcal{B}_i$;
            \State Compute target Q-values for every experience  
            \State ($s_m,a_m,s_{m+1},r_{m+1}$) using the method of 
            \State double DQN:
            \begin{equation}\label{eqn12}
            \begin{array}{l}
            Q_{m} \gets 
            
            \begin{cases}
                 r_{m+1}, \text{ if } \text{terminal state,}\\
                 r_{m+1} + \gamma\cdot Q'(s_{m+1},\\
                 \;\;\;\;\;\;\;\;\;\;\;\arg\max\limits_{a_{m+1}}Q(s_{m+1},a_{m+1}|\theta_i)|\theta'_i), \text{o/w.}

            \end{cases}\nonumber
            \end{array}
            \end{equation}
            \State Calculate Loss function 
            \State $\mathcal{L}(\theta_i)=\mathbb{E}[Q_{m}-$ $Q(s_m,a_m|\theta_i)]^2$,
            \State $\theta_i \gets \theta_i - \alpha \nabla_{\theta_i}\mathcal{L}(\theta_i) $;
            \State /*Update Target Networks*/
            \If {$t\;\%\;r_{update}==0$}
                \State $\theta'_i\gets\theta_i$;
            \EndIf
        \EndFor
    \EndFor
\EndFor
\end{algorithmic}
\end{algorithm}

\subsubsection{Exchange of Complete Local State Information (Level 4)} 
In this level, each UAV shares its complete local state information with other UAVs. At every time step, each UAV will make a decision on its own movement based on the aggregated states of all the UAVs, i.e., the global state information. Accordingly, the number of inputs of each agent's DQN is equal to the dimension of the global state space, and the number of outputs will still be $|\mathcal{A}_i|$, i.e., 5. Compared with Level 3, Level 4 incurs higher computing complexity to each agent during training as the DQN becomes more complicated; sharing the complete state information will also pose the maximum communication overhead. However, it is a more general method that does not need deliberated task-specific reward design as Level 3. In many other problems, precisely extracting the key information to share and well fusing it into the individual reward is non-trivial; whereas level 4 actually provides a worry-free way to obtain a sufficiently good control policy. In terms of the reward function design, Eq. \eqref{eqn10} is reused. 

\subsection{Implementation} \label{subsec.Implementation}
The proposed MA-DQL algorithm is a distributed algorithm where each UAV serves as an agent maintaining its own DQN. The training goes in episodes. Each episode has the same number of time steps as in the considered time horizon. In every step of training, UAVs observe their local states, exchange information, update the DQNs, and make decisions on individual actions in a coordinated manner. The individual policy after training is represented by their respective DQN. Each DQN takes the state information (local states for Level 1-3, and global states for Level 4) as input, and outputs Q values corresponding to every possible individual action. For any given state, the policy takes the action that gives the highest Q value. 

The details of the proposed algorithm are summarized in Algorithm \ref{alg:madqn}. Particularly, target networks are exploited to prevent the target Q values from deviating too much due to a bad experience (Line 4). The method of double Deep Q Networks (DDQN)\cite{van2016deep} is further exploited to reduce overestimation bias of the Q values and make the learning converge more stably (Line 24). Note that as the state and action space are both discrete, the simpler QL can also be technically exploited to solve the problem. We select DQL as it has been proved with considerably better convergence stability than QL\cite{mnih2013playing} due to reduced sample correlation from experience replay. 

\section{Design of Distributed User Connectivity Maximization Algorithm 2 with Heterogeneous Episodes (DUCM-2 Algorithm) }\label{sec.Algorithm2}
This section describes the detailed design of DUCM-2 algorithm with heterogeneous episodes. Different from DUCM-1, DUCM-2 is capable of autonomously handling arbitrary come's and go's of UAVs in the considered time horizon. Moreover, featuring a heterogeneous episode design, the algorithm can obtain the adaptive policy with just one single distributed training.

\subsection{State Space}\label{subsec.statespace2}
In order to handle the dynamic change in the UAV set, it is reasonable to consider that individual UAVs are able to know the real-time (or at least the near real-time) on-off conditions of other UAVs. Here we refer to whether a UAV quits the network as its on-off condition. Denote the on-off condition of UAV $i$ at step $t$ as $L_{i,t}$, and
\begin{equation}\label{live}
 L_{i,t}=\left\{
\begin{array}{rcl}
1,       &      & \text{if UAV $i$ remains in the network}\\
0,       &      & \text{if UAV $i$ quits the network.}
\end{array} \right. 
\end{equation} 
Based on $L_{i,t}$, the on-off conditions of all the UAVs at step $t$ can be monotonically represented as
\begin{equation}\label{livecode}
\mathcal{C}_t = \frac{\sum_{i = 1}^{N^{UAV}_{max}}{L_{i,t}}\cdot 2^{i-1}}{2^{|\mathcal{I}_{t}|}}  ,\;\forall i\in\mathcal{I}_t,\; t = \{1,\cdots,T\}.
\end{equation} 
It can be observed that $\mathcal{C}_t$ is actually a normalized decimal expansion of the binary vector $(L_{1,t},\cdots,L_{i,t},\cdots,L_{N^{UAV}_{max},t})$. The variable $N^{UAV}_{max}$ denotes the maximum possible number of active UAVs in the considered time horizon, which is taken as a preliminary knowledge. In the remainder of the paper, $\mathcal{C}_t$ is referred to as the \textit{Live Code} of the entire UAV set (active or gone), and is included as part of the state space of each UAV. 

The rest of the individual state space reuses the design in Subsection \ref{subsec.statespace}. Thus the individual state of UAV $i$ at step $t$ can be defined as follows, 
\begin{equation}\label{eq.state2}
s_{i,t} = (x_{i,t},\;y_{i,t},\;C_{t},\;t),\;\forall i\in\mathcal{I}_t,\; t = \{1,\cdots,T\},
\end{equation}
where $x_{i,t}$, $y_{i,t}$ represent the coordinates of UAV $i$ at time step $t$. For a DQN design, the input dimension is 4, and the output dimension is the number of possible individual actions, i.e., 5.  

\subsection{Action Space, Information Exchange and Reward Function}\label{subsec.actionspace2}

The action space in Subsection \ref{subsec.actionspace} is reused in this design. For the level of information exchange and the reward function design, it can be found via extensive simulations in Section \ref{sec.Simulation} that Level-3 information exchange and the corresponding reward design generally yield the best convergence performance. Therefore, this algorithm exploits the level-3 information sharing (i.e., sharing UAV positions) and reuses its corresponding reward function design:
\begin{equation}\label{eqnrew}
\begin{split}
&R_i(t) =  \frac{\sum\limits_{u \in \mathcal{U}_t,\; j \in \mathcal{I}_{t}} X_{u,j}(t)}{|\mathcal{I}_{t}|} - \sum_{j \in \mathcal{I}_t, j\ne i}p_{i,j}(t) - f_i, \; \forall{i \in \mathcal{I}_t}\\
&\text{where \;\;\;}  p_{i,j}(t) = max\left(0, \;(1-\frac{d_{i,j}(t)}{2r}) \cdot p_{max}\right), \\
&\;\;\;\;\;\;\;\;\;\;\;\;\;p_{max} = d_p \cdot \frac{|\mathcal{I}_t|}{|\mathcal{U}_t|}, \;\forall{i,j \in \mathcal{I}_t},\text{} i\ne j.
\end{split}
\end{equation}
Please refer to Eq. \eqref{eqn10} and \eqref{eq.penalty} for physical meanings of the symbols.

\subsection{Implementation} \label{subsec.Imple}
The following implementations are realized in order to achieve an adaptive UAV relocation strategy \textit{i)} that can handle arbitrary UAV quit's and join-in's in the considered time horizon, ii) in just a single distributed training, and \textit{iii)} with increased training efficiency compared to the DUCM-1 algorithm.

\subsubsection{Heterogeneous episodic training}
Two types of episodes are designed which alternate along the training. In odd number of episodes, full set of UAVs stay alive for the entire considered time horizon; while in even number of episodes, UAVs randomly and sequentially quit the network from a full set to a single one every certain time steps. The reasons that a full set of UAVs are trained in an entire odd-indexed episode are as follows. Training for a full set of UAVs is more complex than that for a subset of UAVs, thus requiring more experiences collected and training iterations. In addition, the optimal positions of the full set can serve as a good reference for the optimal positioning of a subset of UAVs, thus facilitating the entire convergence.

In order to make the derived strategy generalized to handle quits and join-ins of arbitrary UAVs, in each even number of episode, the order of UAV quits are totally random. The interval between two consecutive quits is set to $2(M-1)$, which is the minimum number of steps between the two farthest possible UAV positions in the target region.


\subsubsection{Parallel complementary training for UAV quits and join-ins} As the derived strategy is expected to handle both UAV quits and join-ins, episodes for sequentially increasing the number of UAVs should have been considered. However, in order to improve the time efficiency of training, a complementary training mechanism is implemented in even number of episodes. We consider two independent environment copies in each even number of episode. The two copies are exactly the same except the randomness, making them mutually independent. When a UAV quits the network in the first copy, it automatically enters the other copy where it continues to be trained with any existing UAVs for the same optimization goals. The set of UAVs in the two copies are complementary to each other and the training in the two copies take place at the same time. For each UAV, the collected experiences in either copy will be combined into its single experience buffer to update its Q network. Accordingly, the on-off condition of one UAV in the second environment copy, denoted as $\mathcal{D}_{i,t}$, is calculated as
\begin{equation}\label{destroy}
\mathcal{D}_{i,t} =  \sim  {L}_{i,t}, \;\forall i\in\mathcal{I}_t,\; t = \{1,\cdots,T\}.
\end{equation} 
The live code of each UAV in copy 2 will be 
\begin{equation}\label{destroycode}
\mathcal{C}_{t,2} = \frac{\sum_{i = 1}^{N^{UAV}_{max}}{\mathcal{D}_{i,t}}\cdot 2^{i-1}}{2^{|\mathcal{I}_{t}|}} = \overline{\mathcal{C}_t}  ,\;\forall i\in\mathcal{I}_t,\; t = \{1,\cdots,T\}.
\end{equation} 
which is actually the 1's complement of $\mathcal{C}_t$. The same state, action and reward designs in \eqref{eqnrew} are used but in the context of the second environment copy. The detailed implementation of DUCM-2 algorithm is described in Algorithm \ref{alg:madqn1}.

\begin{algorithm}[!ht]
\caption{Distributed User Connectivity Maximization Algorithm 2 (DUCM-2)}\label{alg:madqn1}
\begin{algorithmic}[1]
\ForAll{$i \in \mathcal{I}_t$}
    \State Randomly initialize the main and target DQNs: 
    \State $Q(\mathcal{S}_i,\mathcal{A}_i|\theta_i$), $Q'(\mathcal{S}_i,\mathcal{A}_i|\theta'_i)$ = $Q(\mathcal{S}_i,\mathcal{A}_i|\theta_i)$;
    \State Initialize experience replay buffer $\mathcal{B}_i$;
\EndFor
\State Initialize two identical environment copies with different random seeds: $env1, env2$;
\For{episode := 1:$N_{Eps}$}
    \State {Reset all UAVs to initial states $s_{i,0},\forall i\in\mathcal{I}_t$;}

    \For{time step $t$ := $1:T$}
        \If {mod(episode,\;2) == 0}
        \If {mod($t$,\;$2 \cdot (M-1)$) == 0}
            \State {One remaining UAV in $env1$ quits and }
            \State enters $env2$; 
        \EndIf
        \State{Update $\mathcal{C}_t$ and $\overline{\mathcal{C}_t}$;}
    \EndIf
        \ForAll{$i \in \mathcal{I}_t$}
            \State {$env = (L_{i,t}==1)\mathord{?}env1 \mathord{:} env2$;}
            \State {/*$\epsilon$-greedy action selection*/}
            \State $\beta\gets$ Rand(0,1); 
            \If {$\beta\le\epsilon$}
                \State {$a_{i,t} \gets$ Rand([0,1,2,3,4]);}
            \Else
                \State {$a_{i,t} \gets \arg\max_{a_{i}} Q(s_{i,t},a_i|\theta_i)$;}
            \EndIf
            \State {Execute $a_{i,t}$ and observe the next state $s_{i,t+1}$;}
            \State {Share information with other UAVs in $env$ }
            \State according to Level 3;
            \State {Calculate reward $R_i(t+1)$ according to Eq. }
            \State \eqref{eqnrew};
            \State {Store the experience $(s_{i,t},a_{i,t},s_{i,t+1},R_i(t+1)$) }
            \State into Buffer $\mathcal{B}_i$;
        
            \State {/*Update of DQNs*/}
            \State Randomly sample a mini-batch from $\mathcal{B}_i$;
            \State Compute target Q-values for every experience
            \State ($s_m,a_m,s_{m+1},r_{m+1}$) using the  method of 
            \State double DQN: 
            \begin{equation}
            \begin{array}{l}
            Q_{m} \gets 
            
            \begin{cases}
                 r_{m+1}, \text{ if } \text{terminal state,}\\
                 r_{m+1} + \gamma\cdot Q'(s_{m+1},\\
                 \;\;\;\;\;\;\;\;\;\;arg\max\limits_{a_{m+1}}Q(s_{m+1},a_{m+1}|\theta_i)|\theta'_i), \text{o/w.}
            \end{cases}\nonumber
            \end{array}
            \end{equation}
            \State Calculate Loss function 
            \State $\mathcal{L}(\theta_i)=\mathbb{E}[Q_{m}-Q(s_m,a_m|\theta_i)]^2$,
            \State $\theta_i \gets \theta_i - \alpha \nabla_{\theta_i}\mathcal{L}(\theta_i) $;
            \State /*Update Target Networks*/
            \If {$t\;\%\;r_{update}==0$}
                \State $\theta'_i\gets\theta_i$;
            \EndIf
        \EndFor
    \EndFor
\EndFor
\end{algorithmic}
\end{algorithm}

\section{Numerical Results}\label{sec.Simulation}
\subsection{Simulation Setup}
We consider an area of 10x10 square units with each unit being 100 meters. The environment is implemented in Python utilizing the Gym toolkit by OpenAI. PyTorch library is employed to develop the proposed DUCM-1 and DUCM-2 algorithms. 
Mean square error is adopted as the loss function for DQN training and gradient clipping is implemented to avoid the issues of exploding and vanishing gradients. The training is conducted on a Windows 10 mobile AI workstation with Intel Core i7-11850H CPU @ 2.50GHz, 64GB RAM, and NVIDIA RTX A5000 GPU. The training has maximum 1000 episodes. There are 100 time steps for DUCM-1 algorithm while 150 time steps for DUCM-2 algorithm. 
The main simulation parameters are summarized in Table \ref{tab:simparam}. Note that in DUCM-1, DQN for L4 information exchange is more complicated than other levels due to the consideration of the global states. DQN of DUCM-2 is more complicated than DUCM-1 as it considers a highly dynamic environment where UAVs can arbitrarily quit or join in the network.

\begin{table}[!ht]
\footnotesize
\centering
\renewcommand{\arraystretch}{1}

\begin{tabular}{!{\vrule width0.8pt}l|l!{\vrule width0.8pt}}\Xhline{0.8bp}
\multicolumn{1}{!{\vrule width0.8pt}c|}{\gape{\bfseries Parameters}} & \multicolumn{1}{c!{\vrule width0.8pt}}{\gape{\bfseries Values}} \\ 
         \hline
         \rowcolor[gray]{0.9}
         Number of UAVs and Users & (5, 100)\\
         Number of RBs per UAV & {20} \\
         \rowcolor[gray]{0.9}
         {RB bandwidth $BW_{rb}$} & {180 KHz} \\
         {Percentage $p$ of being hotspot users} & {0.8} \\
         \rowcolor[gray]{0.9}
         {Altitude $H$} and aperture angle $\theta$ & {$350m$ and $60^{\circ}$} \\
         Spectrum center frequency $f_c$ & 2GHz\\
         \rowcolor[gray]{0.9}
         Transmit and noise psd: ($P_t, n_0$)& (-49.5dBm, -174dBm)\\ 
         Min. user throughput $r_{min}$ & 250kbps\\
         \rowcolor[gray]{0.9}
         LoS path loss parameter $\eta$ & 1dB\\
         {Out-of-boundary penalty,} $f_i$ & {2 if out, 0 otherwise} \\
         \rowcolor[gray]{0.9}
         {Constant weighting factor, $d_p$} & {0.25} \\
         {Learning rate of DQN, $\alpha$} & {DUCM-1: $2.5e$-4, } \\
                                          & {DUCM-2: $3.5e$-4}\\
         \rowcolor[gray]{0.9}
         {Replay batch size} & {512} \\
         {Discount factor $\gamma$} & {0.95} \\
         \rowcolor[gray]{0.9}
         {DQN structure} & {\scriptsize{Hidden nodes: 2x400 (DUCM-1,}}\\
         \rowcolor[gray]{0.9}
         & {\scriptsize{L1, L2, L3), 3x256 (DUCM-1, L4),}}\\
         \rowcolor[gray]{0.9}
         & {\scriptsize{3x400 (DUCM-2, L3). A norm}}\\
         \rowcolor[gray]{0.9}
         & \scriptsize{layer comes after each hidden layer}\\
         {Epsilon $\epsilon$} & {0.1} \\    
         \rowcolor[gray]{0.9}
         {Target network updating rate}& {1/10} \\
\hline
\end{tabular}
\caption{Simulation Parameters}
\label{tab:simparam}
\end{table}

\subsection{Simulation Results for DUCM-1 Algorithm}\label{subsec.simuresults}
\begin{figure}[!ht]
\centering
\includegraphics[width=.65\linewidth]{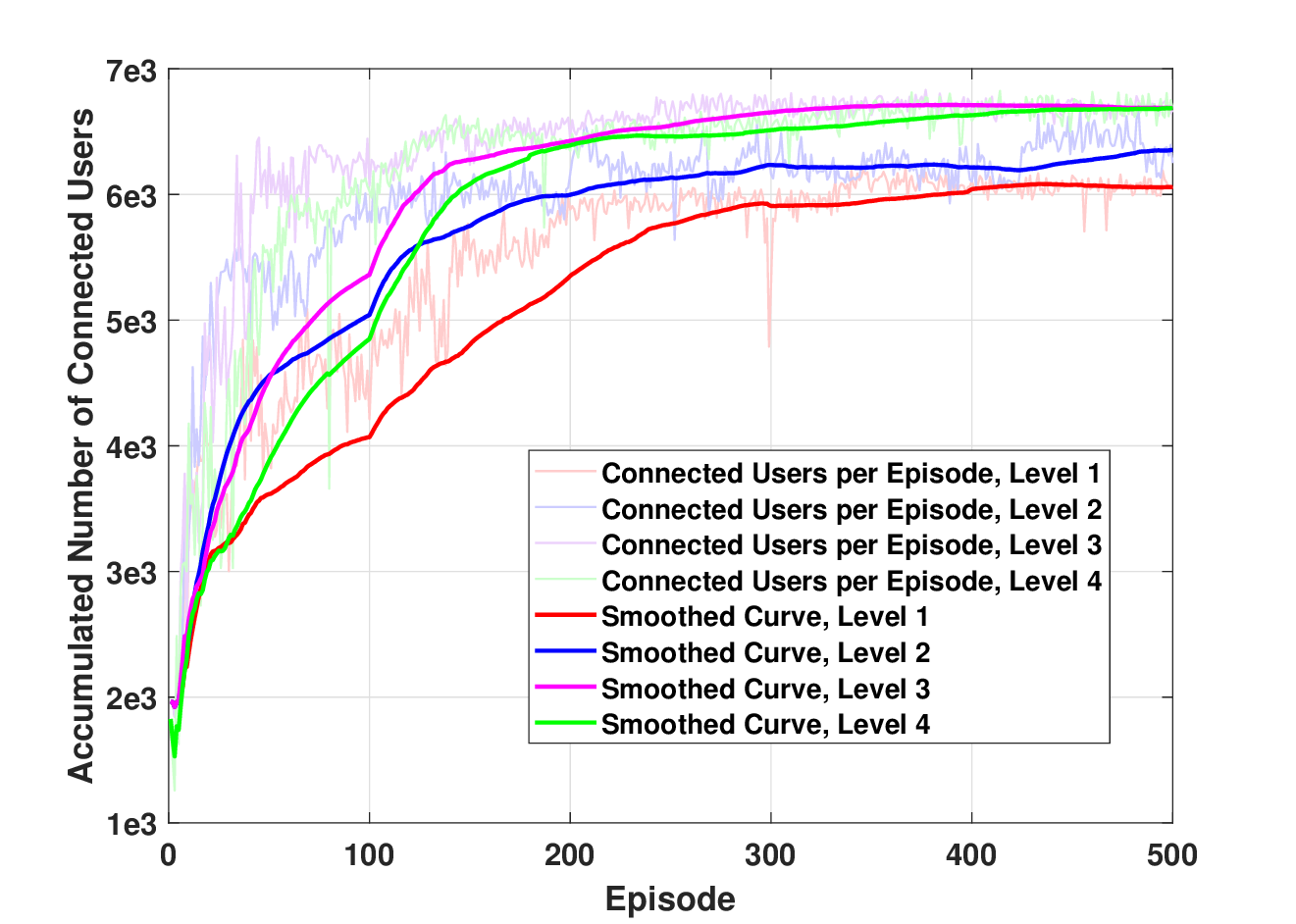}
\caption{Convergence of the number of connected users per episode for different levels of information exchange.}
\label{fig:run002}
\end{figure}

\begin{figure}[!ht]
\centering
\includegraphics[height=0.38\linewidth, width=.45\linewidth]{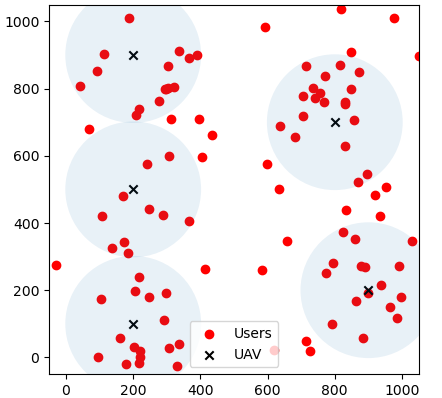}
\caption{Final UAV positions and user connectivity under Level 3 and Level 4}
\label{fig:bestcoverage}
\end{figure}

We first compare the convergence performance of different information sharing levels. Default simulation parameters in Table \ref{tab:simparam} are used. The results are shown in Fig. \ref{fig:run002}. The accumulated numbers of served users in the time horizon $T$ are compared instead of the episodic rewards. This is because different levels have different reward function designs and thus a direct reward comparison cannot reflect the true user connectivity performance. It can be observed that Level 1 yields fastest convergence yet the worst user connectivity performance. Level 2 converges the most slowly with a suboptimal policy. Level 3 and level 4 achieve almost the same and the best converged value, yet level 3 converges faster than level 4. The reason is as follows. Level 1 exchanges the least amount of information. Each drone makes decisions relying solely on its local observations. Level 2 exchanges individual reward information to derive the global user connectivity thus achieving better performance than level 1. However, each drone lacks clear guidance on how to move to improve its own reward. Level 4 has the similar issue. But instead of the consequential global user connectivity, each drone gets to know the global states which embraces more information to achieve better convergence. The downside is that the global state space has more dimensions which require more explorations to find the optimal policy, i.e., slower convergence. Level 3 leverages task-specific knowledge that a more dispersed UAV distribution may potentially result in a better overall user connectivity. This knowledge guides the UAVs to move away from others, thus achieving the best convergence. Fig.\ref{fig:bestcoverage} shows the final user connectivity conditions obtained under Level 3 and Level 4, which are the same.

\begin{figure}[!ht]
\centering
\includegraphics[width=.65\linewidth]{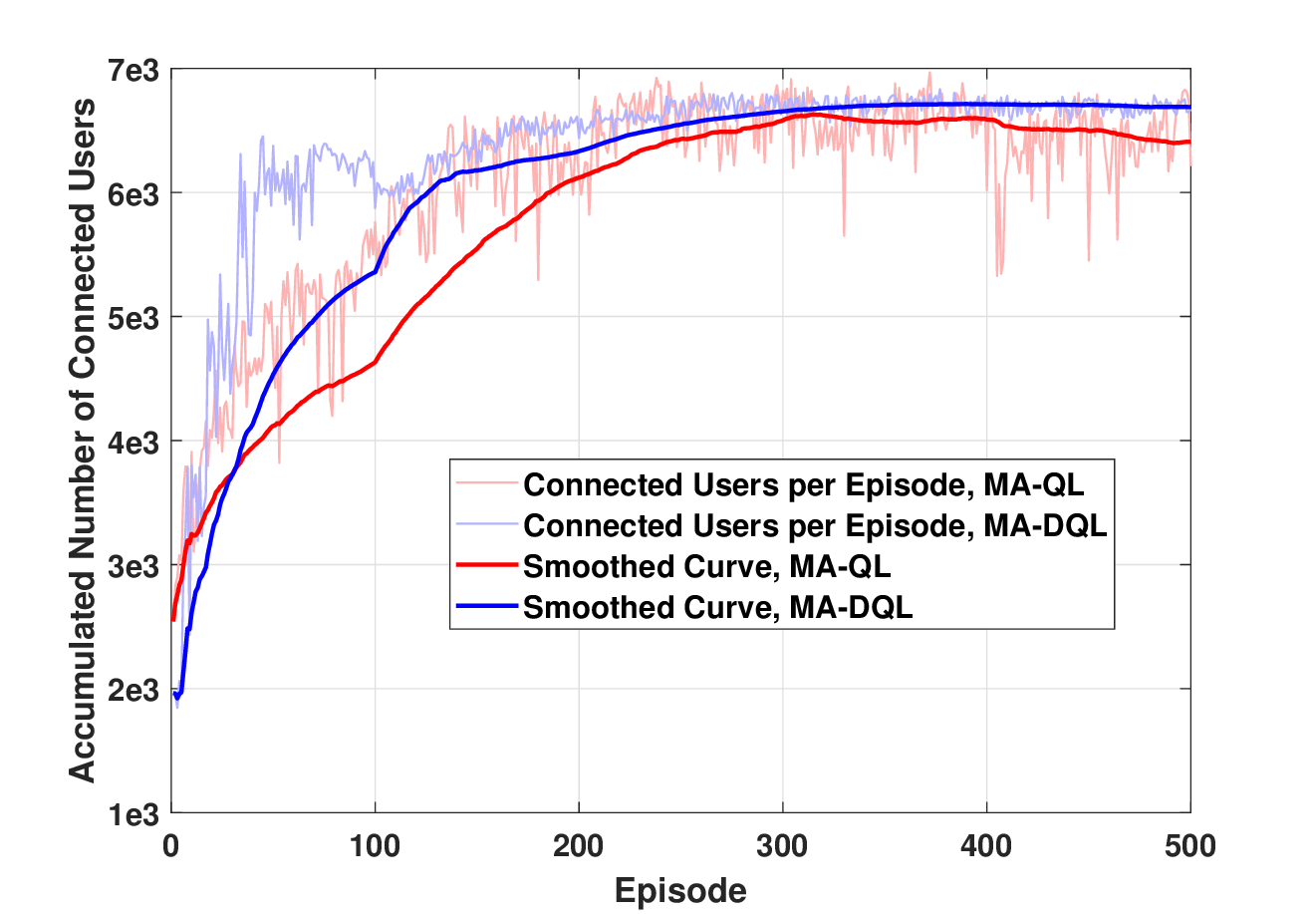}
\caption{Convergence comparison between MA-QL and MA-DQL.}
\label{fig:compare}
\end{figure}

We further compare the convergence performance between MA-QL and MA-DQL under Level 3 information sharing in Fig. \ref{fig:compare}. It can be seen that MA-QL and MA-DQL achieve similar converged values, but MA-DQL yields considerably faster and more stable convergence (i.e., much less fluctuation in the flat region). This is due to the DQL's advantages in implementing experience replay and target networks.

\begin{figure}[!ht]
\centering
\includegraphics[width=.70\linewidth]{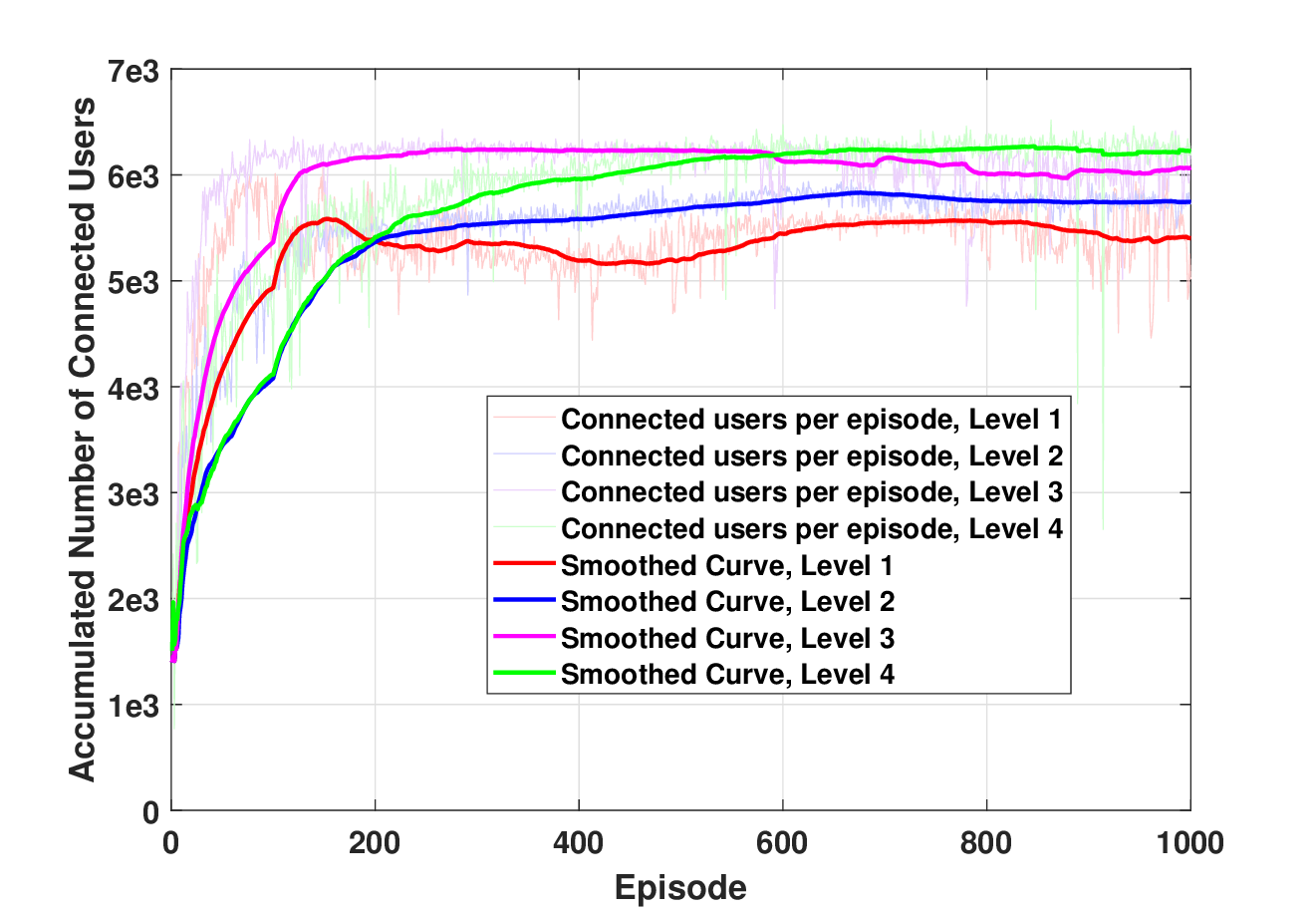}
\caption{Convergence of different levels with dynamic user distribution.}
\label{fig:dynamic}
\end{figure}

\begin{figure}[!ht]
\centering
\subfigure[First half] {\includegraphics[width=.43\linewidth, height=1.6in]{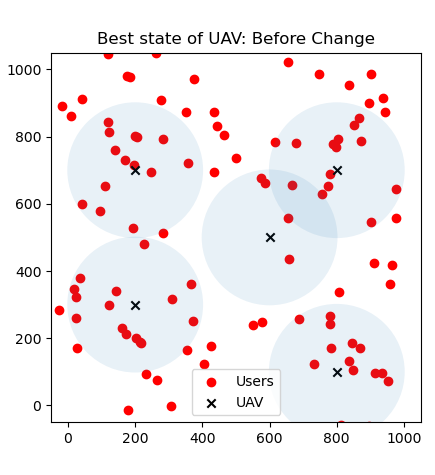}\label{Fig.1}} 
\subfigure[Second half] {\includegraphics[width=.43\linewidth, height=1.6in]{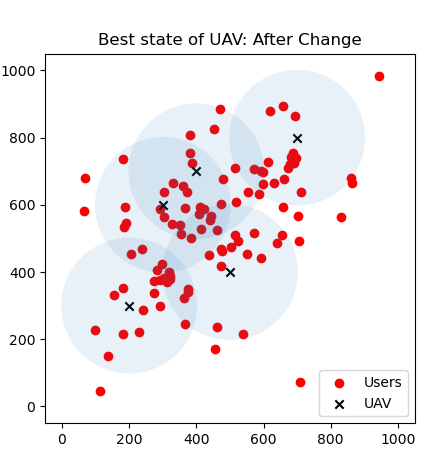}\label{Fig.2}}
\caption{Final UAV positions and user coverage with Level 3 information exchange: Dynamic user distribution}\label{fig:dynamic1}
\end{figure}

Lastly, we execute DUCM-1 algorithm under different levels with a dynamic user distribution. A simple case is considered where the users are more dispersed in the first half of the time horizon and then become more concentrated in the second half. Fig. \ref{fig:dynamic} shows the convergence performance. It can be found that among all the levels, level 3 converges the fastest but the curve declines slightly after convergence is achieved. Level 4 converges more slowly but yields slightly better converged values than level 3. The best achieved user connectivity conditions are presented in Fig. \ref{fig:dynamic1}.

\vspace{-0.2in}

\subsection{Simulation Results for DUCM-2 Algorithm}

\begin{figure}[!ht]
\centering
\includegraphics[width=.65\linewidth]{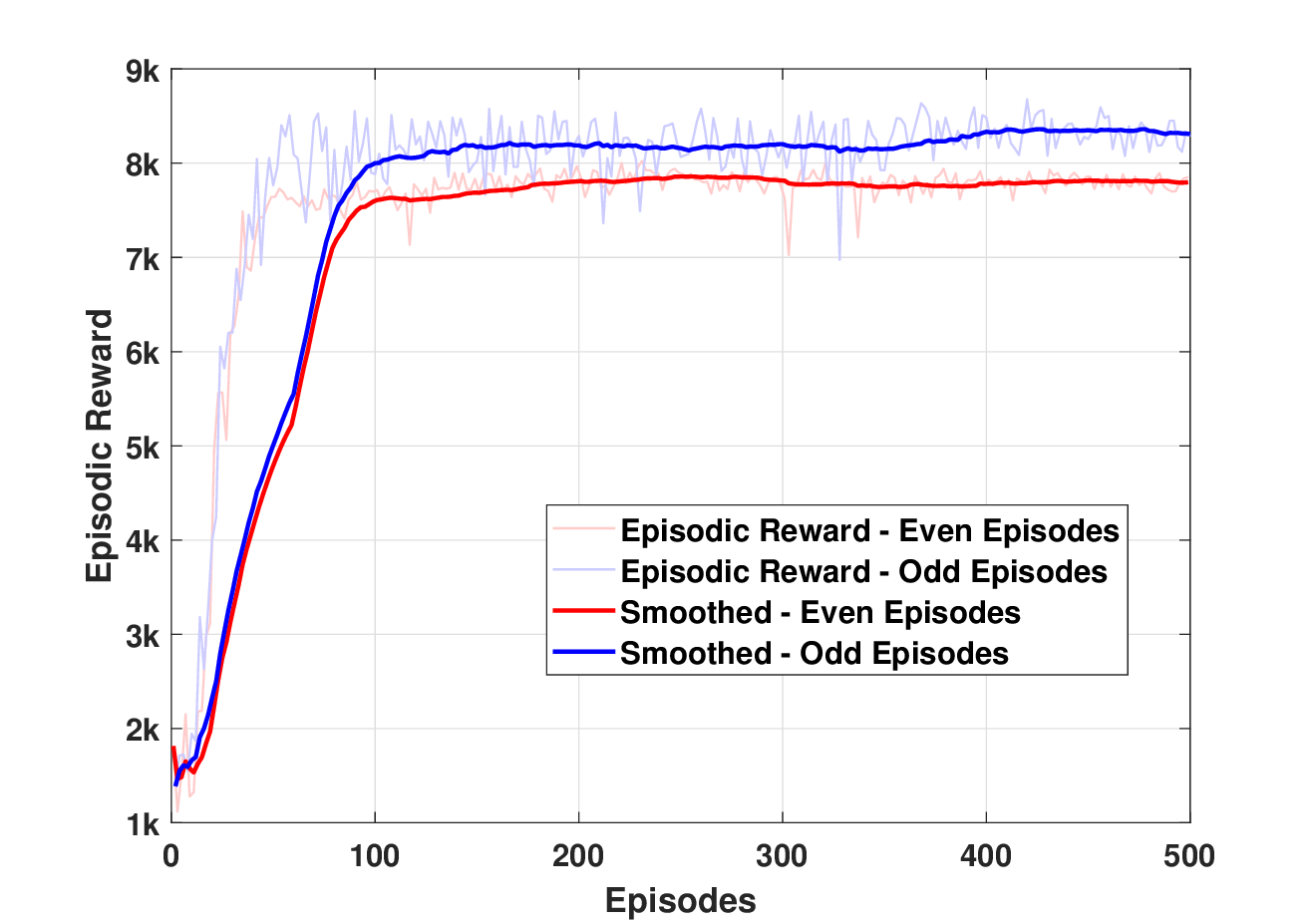}
\caption{Convergence performance of DUCM-2 algorithm.}
\label{fig:convergence_DUCM2}
\end{figure}

The performance of DUCM-2 algorithm is then evaluated. As level-3 information exchange has been shown to yield the best convergence, all the following experiments are carried out under level 3. The convergence performance of DUCM-2 algorithm is provided in Fig. \ref{fig:convergence_DUCM2}. Due to the heterogeneity of the episodes, there are 2 curves for the even and odd number of episodes, respectively. The convergence is obtained faster than that of DUCM-1 because the DQNs are more complex (thus learning faster) and use a larger learning rate (i.e., $3.5e$-3 vs. $2.5e$-3). 

After the optimal distributed policy is obtained, simulations are conducted to test its capabilities of handling arbitrary UAV quits and join-ins. Three types of dynamic UAV change are tested: $i$) the number of UAVs start from 5 and 4 UAVs quit the network sequentially in a random order, $ii$) the number of UAVs start from 1, and 4 UAVs join in the network sequentially, and $iii$) the number of UAVs start from 3 and the next 4 events of change are random quits or join-ins (with the number of active UAVs kept between 1 and 5). We test all the possible cases of type $i$) (120 different orders of quits) and type $iii$ (12 different orders as the number of active UAVs needs to be between 1 and 5). Only 1 order is considered for type $ii$) as UAVs are identical.

\begin{figure}[!ht]
\centering
\subfigure[Random sequential UAV quits] {\includegraphics[width=.7\linewidth]{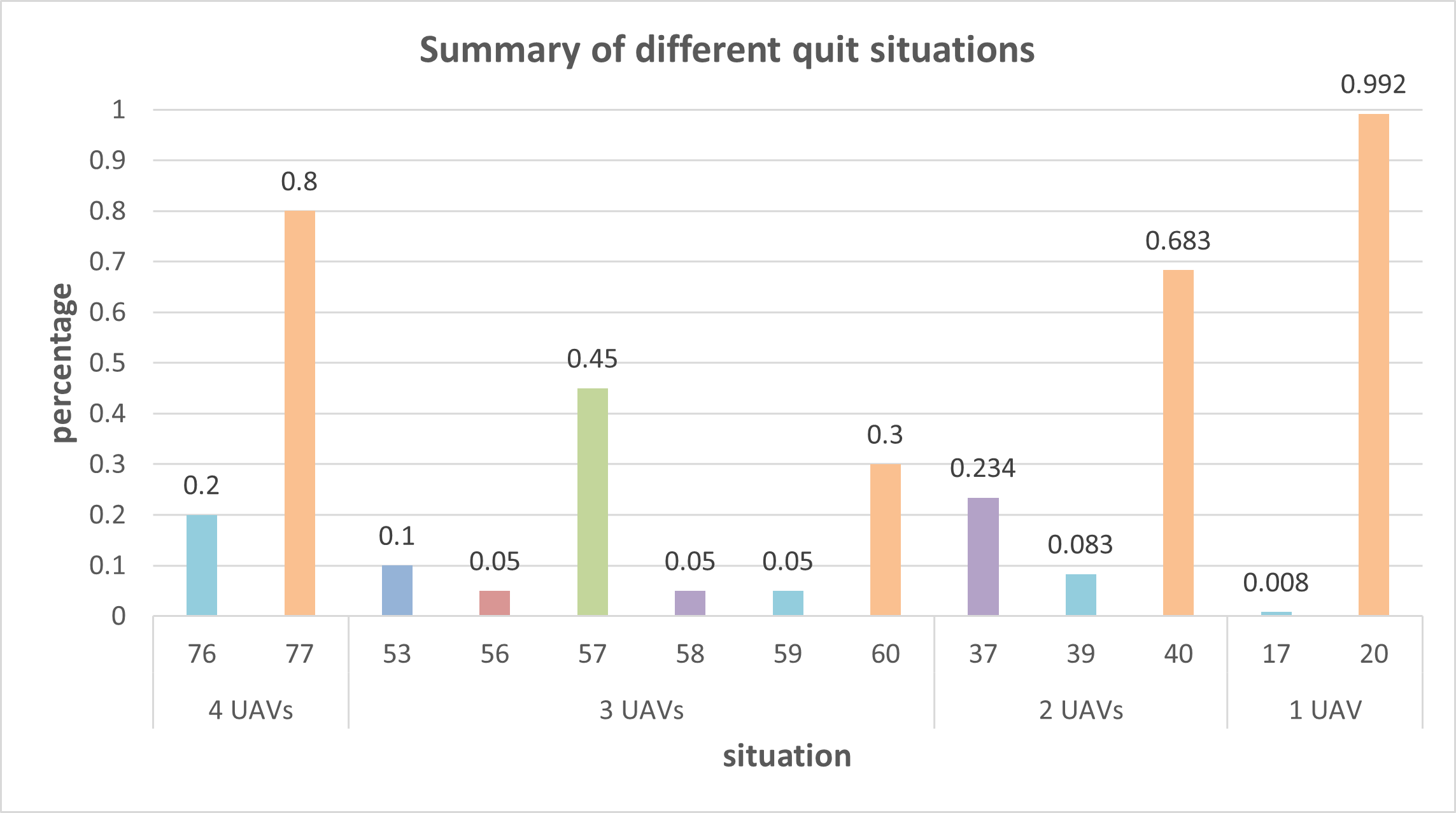}\label{Fig.test1}}\\
\subfigure[Random sequential quits and join-in's] {\includegraphics[width=.7\linewidth]{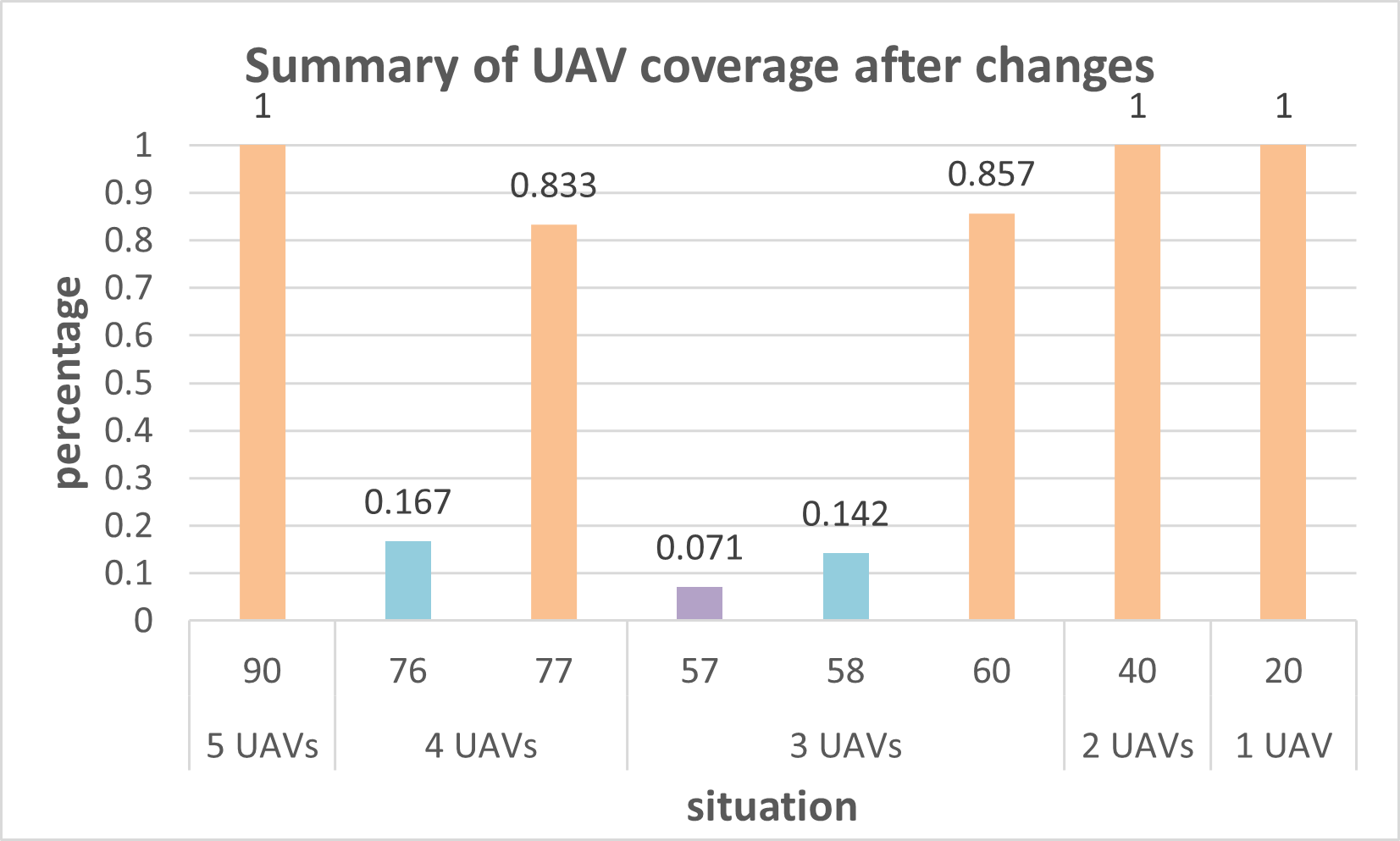}\label{Fig.test2}}
\caption{Exhaustive test results of the obtained distributed user connectivity maximization strategy under DUCM-2 algorithm.}\label{Fig.testresults}
\end{figure}

\begin{figure}[!ht]
\centering
\subfigure[Full set] {\includegraphics[width=.45\linewidth]{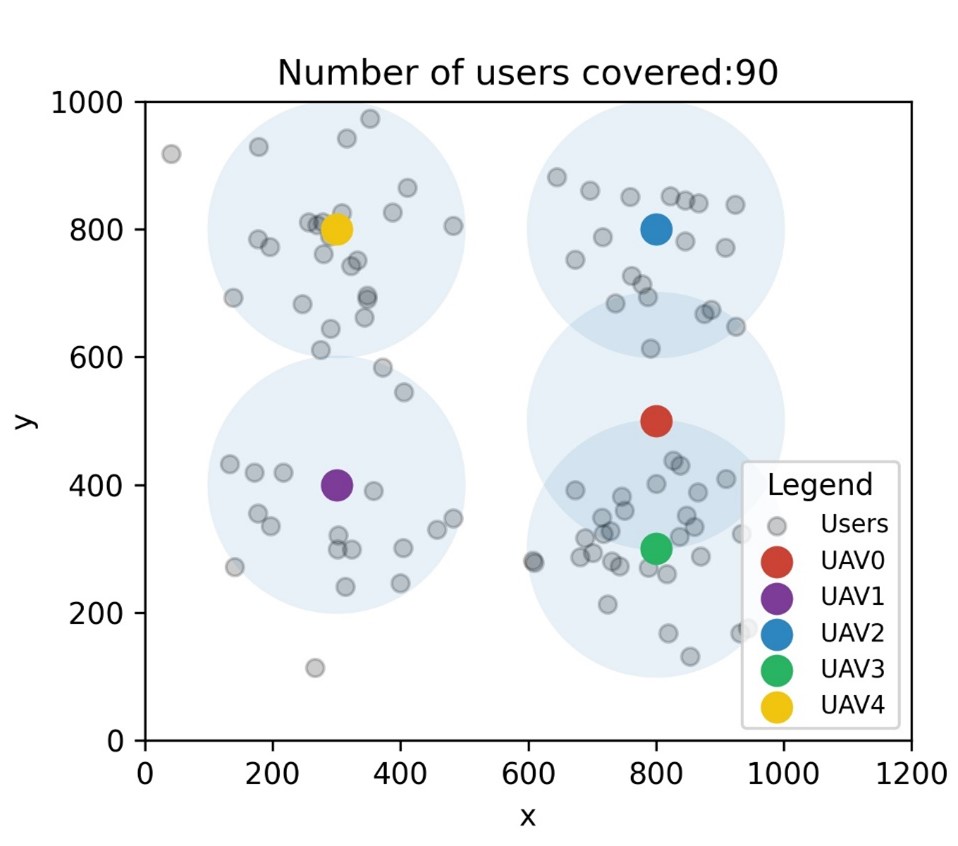}\label{Fig.des2-1}} 
\subfigure[UAV 1 quits] {\includegraphics[width=.45\linewidth]{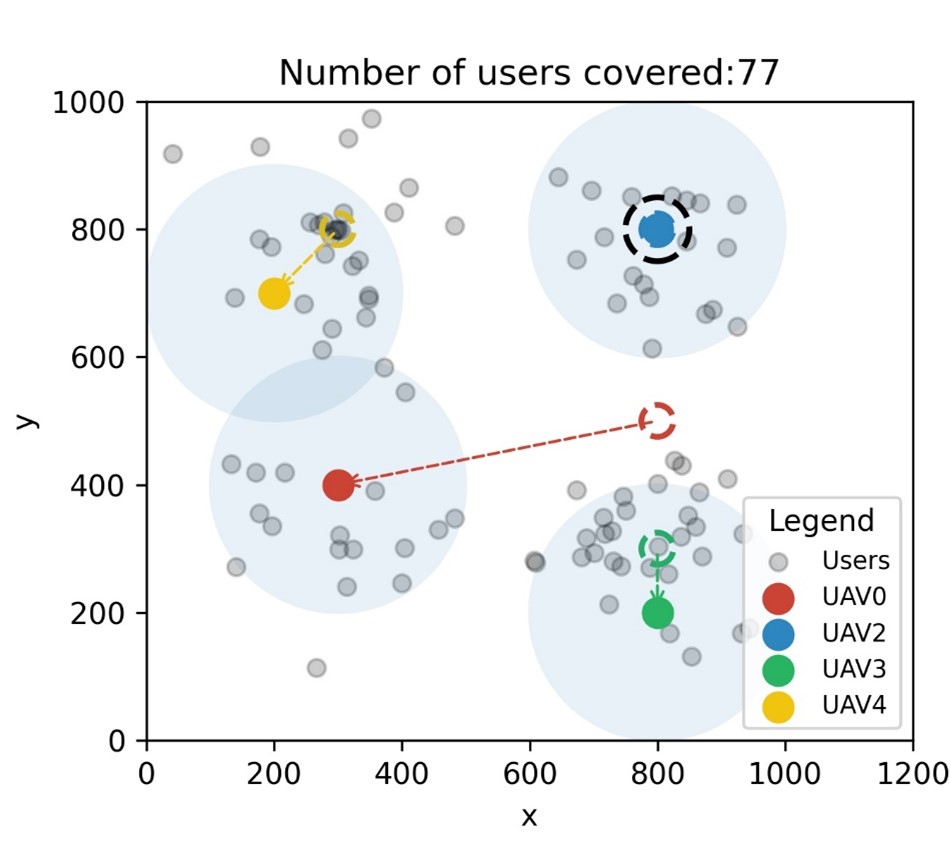}\label{Fig.des2-2}}
\subfigure[UAV 0 quits] {\includegraphics[width=.45\linewidth]{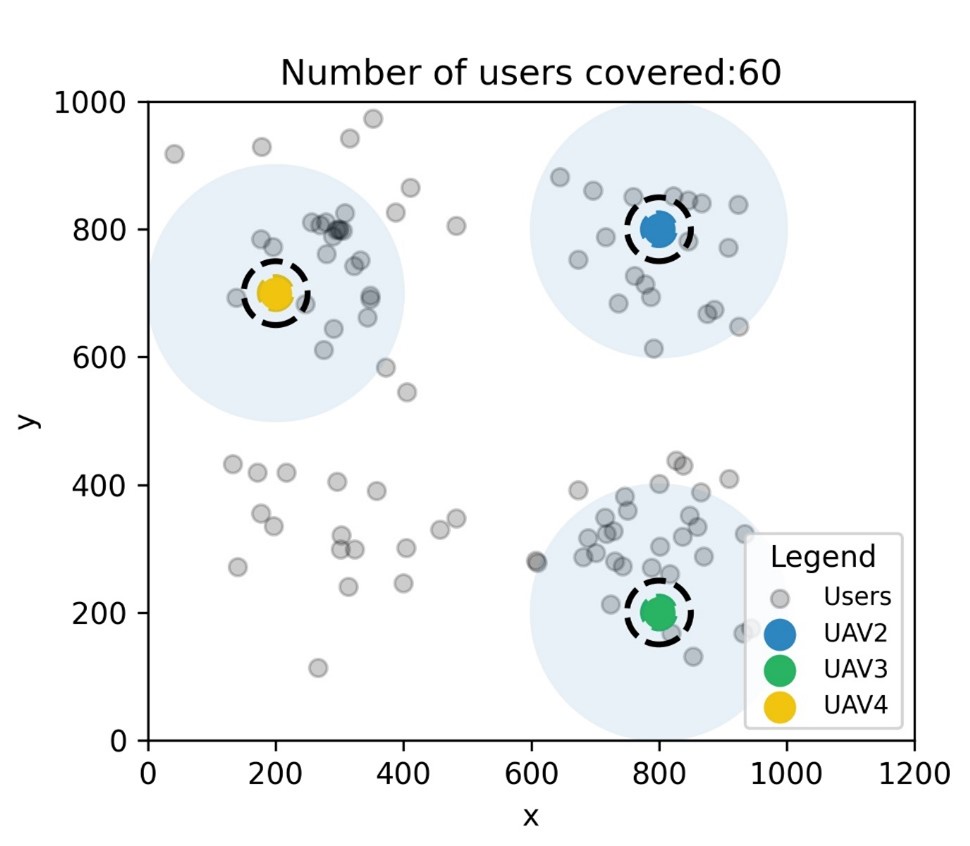}\label{Fig.des2-3}}
\subfigure[UAV 2 quits] {\includegraphics[width=.45\linewidth]{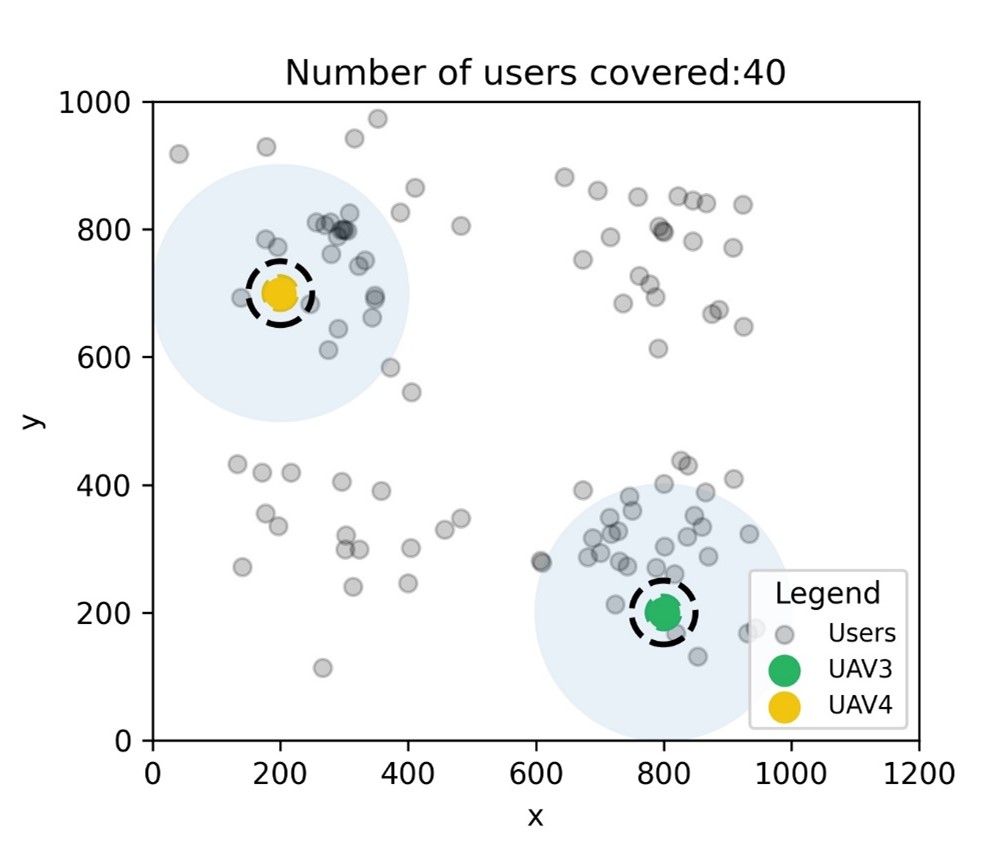}\label{Fig.des2-4}}
\subfigure[UAV 3 quits] {\includegraphics[width=.45\linewidth]{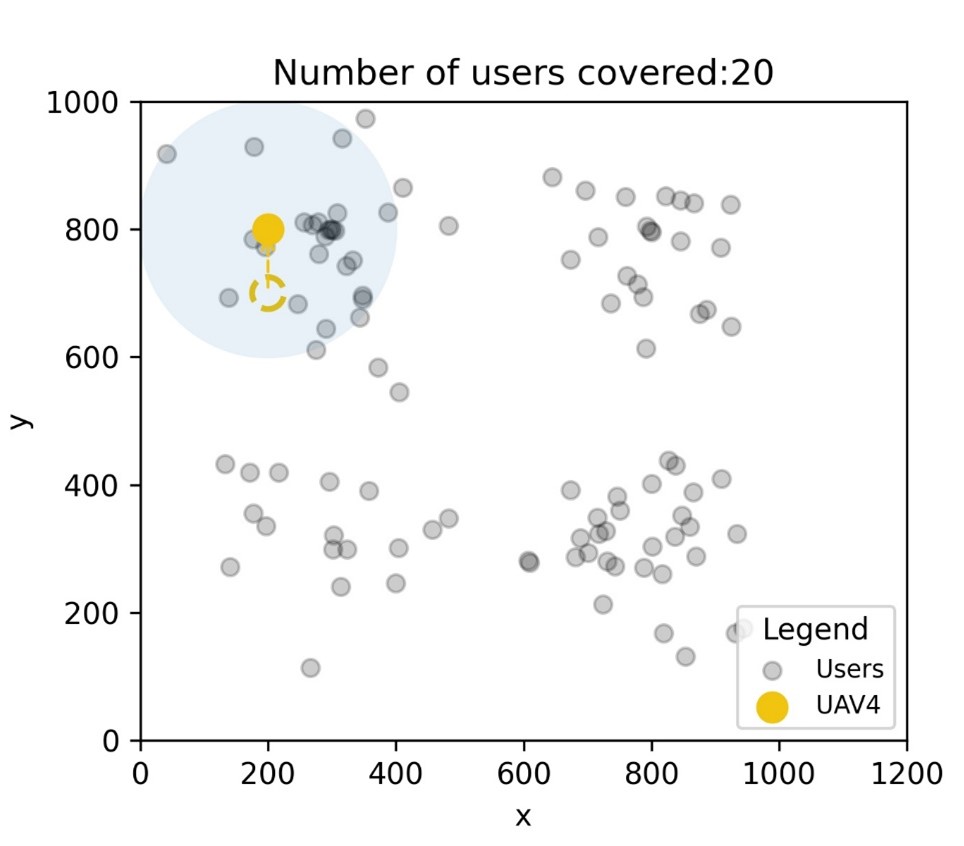}\label{Fig.des2-5}}
\caption{Optimal coverage of active UAVs when UAVs quit in an index order of $\{$1,0,2,3$\}$}\label{fig:des2}
\end{figure}

\begin{figure}[!ht]
\centering
\subfigure[Starts with UAV 4] {\includegraphics[width=.45\linewidth]{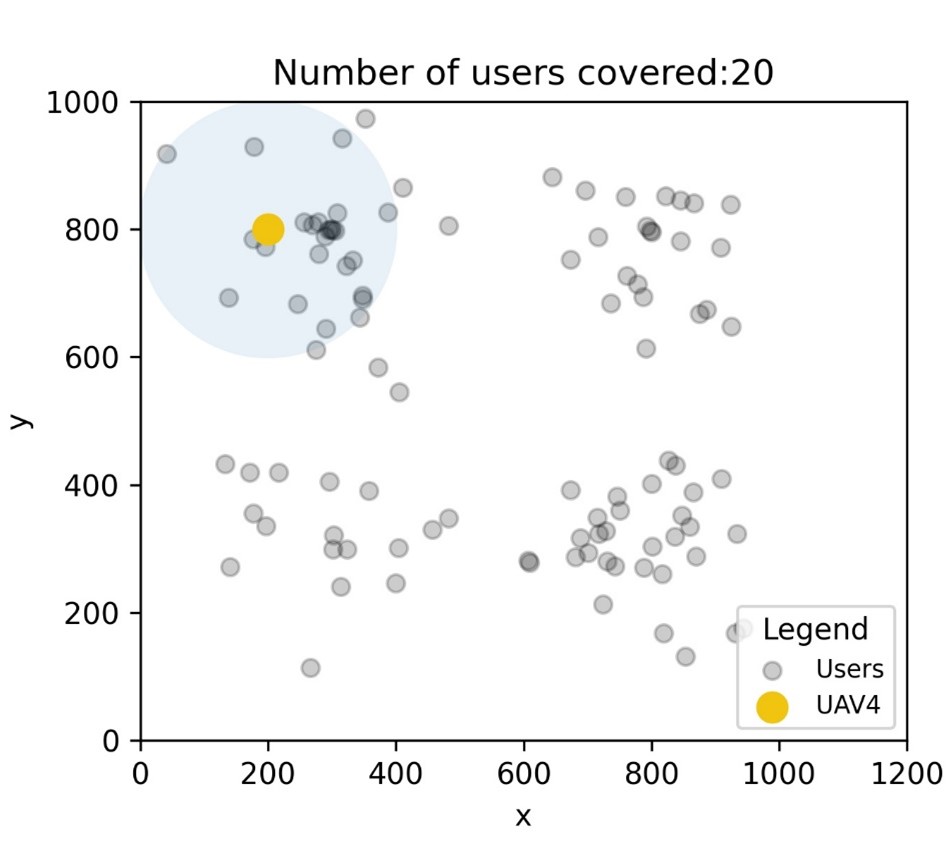}\label{Fig.fly3-1}} 
\subfigure[UAV 3 joins in] {\includegraphics[width=.45\linewidth]{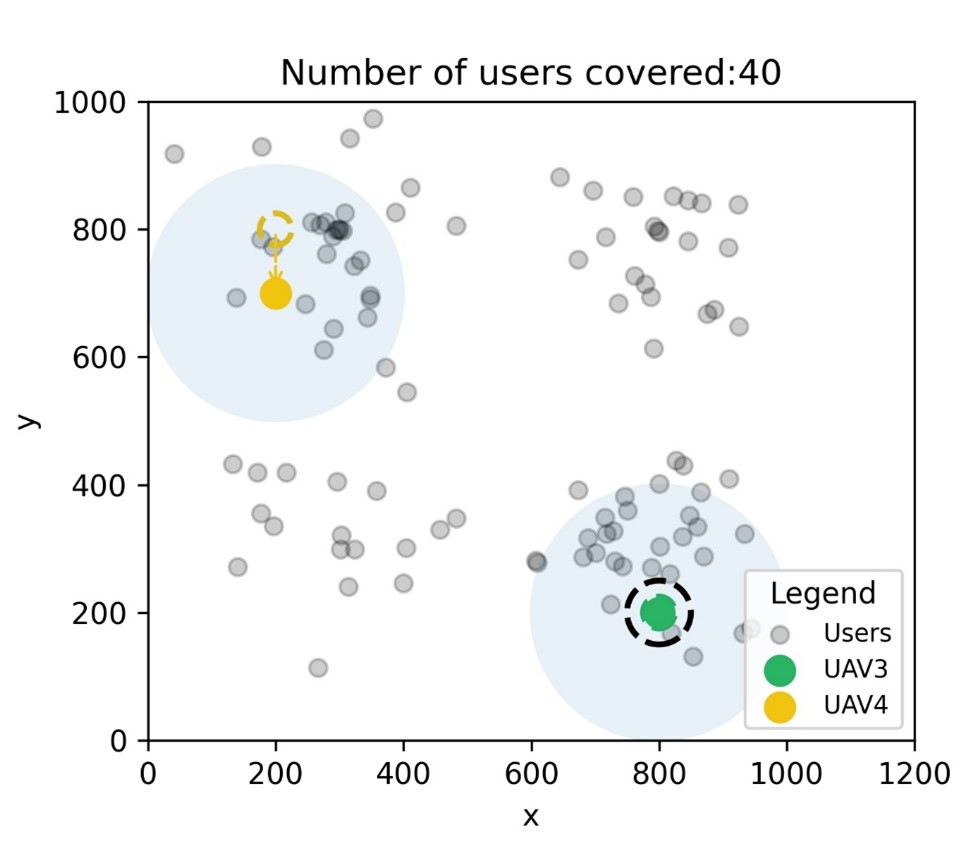}\label{Fig.fly3-2}}
\subfigure[UAV 0 joins in] {\includegraphics[width=.45\linewidth]{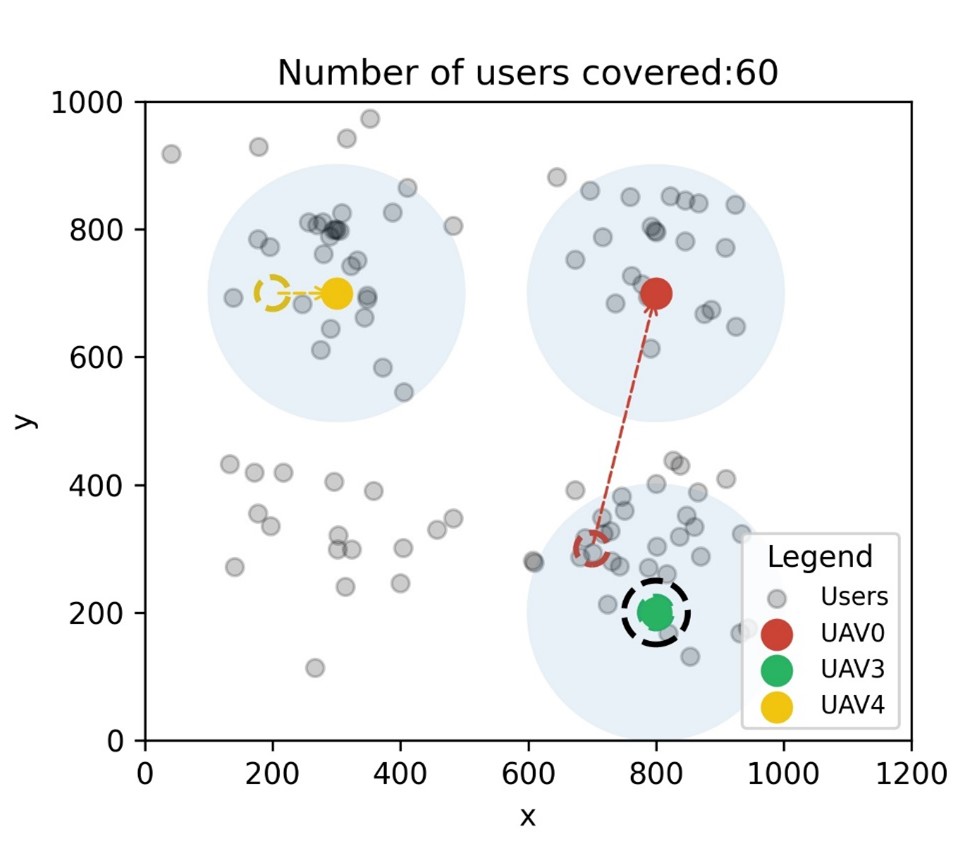}\label{Fig.fly3-3}}
\subfigure[UAV 2 joins in] {\includegraphics[width=.45\linewidth]{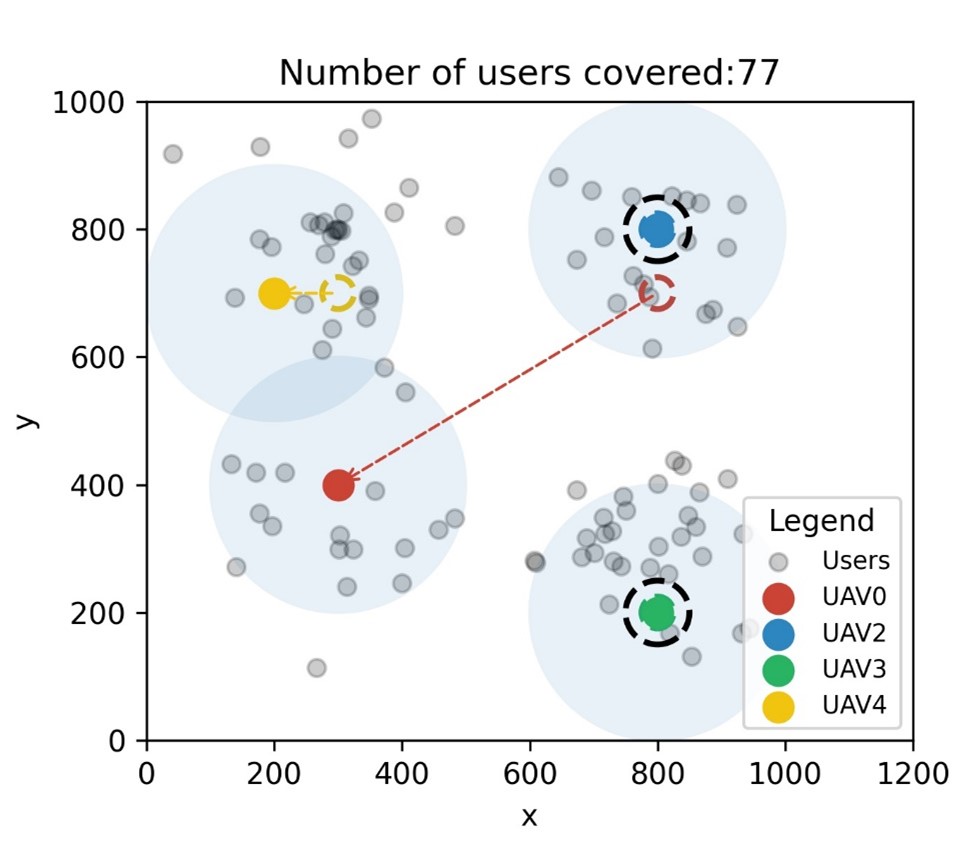}\label{Fig.fly3-4}}
\subfigure[UAV 1 joins in] {\includegraphics[width=.45\linewidth]{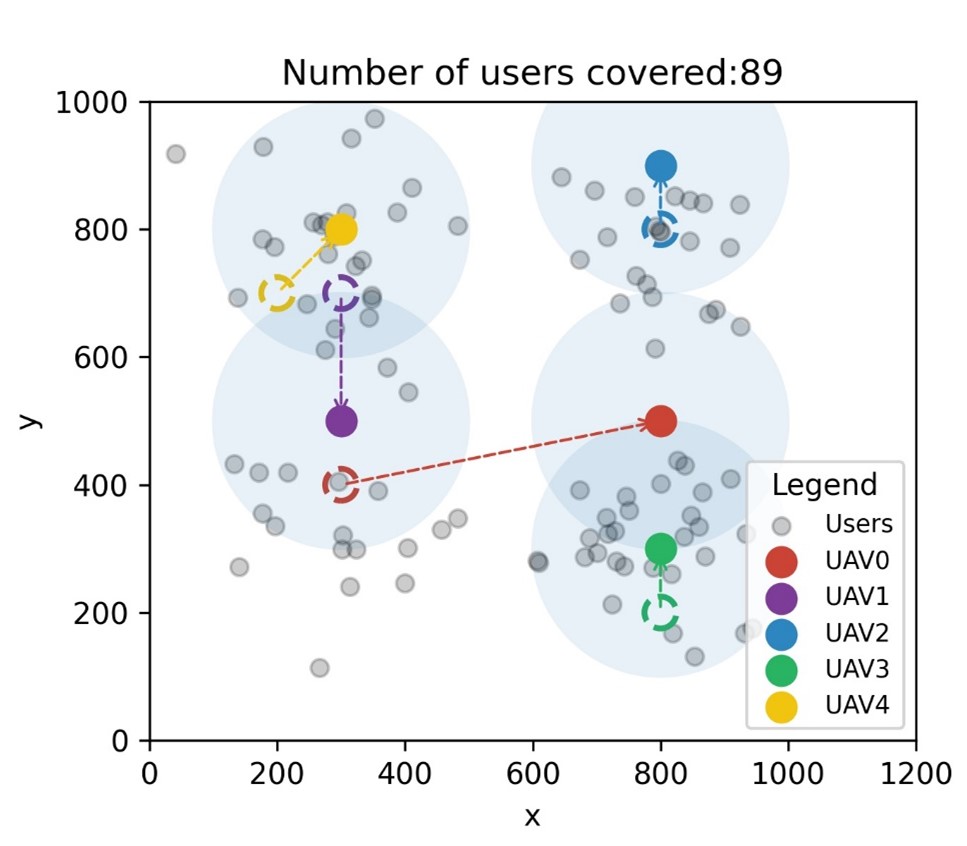}\label{Fig.fly3-5}}
\caption{Optimal coverage of active UAVs when UAVs join in in an index order of $\{$4,3,0,2,1$\}$}\label{fig:fly3}
\end{figure}

\begin{figure}[!ht]
\centering
\subfigure[Start with 3 UAVs] {\includegraphics[width=.45\linewidth]{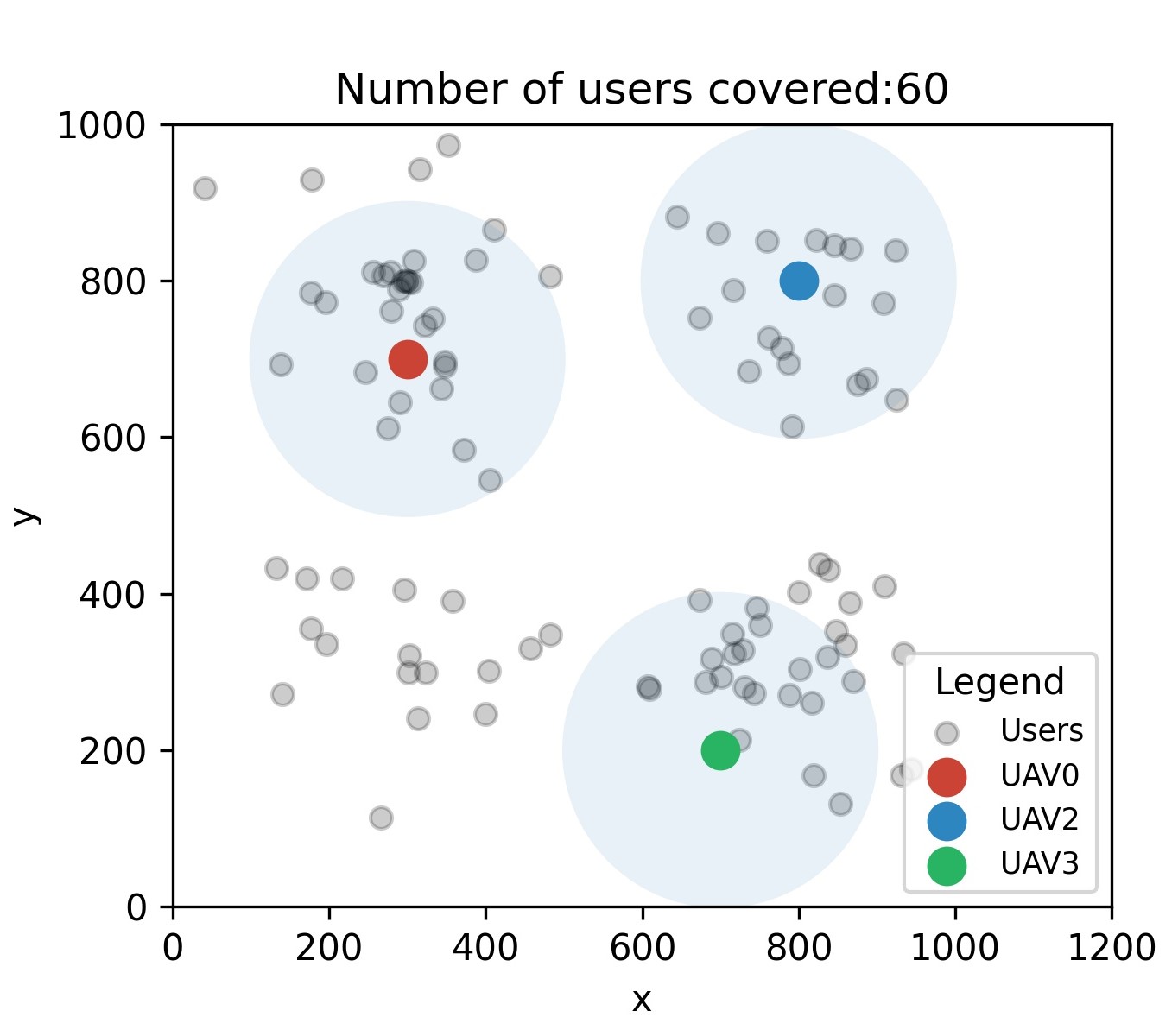}\label{Fig.fly3-11}} 
\subfigure[1 UAV joins in] {\includegraphics[width=.45\linewidth]{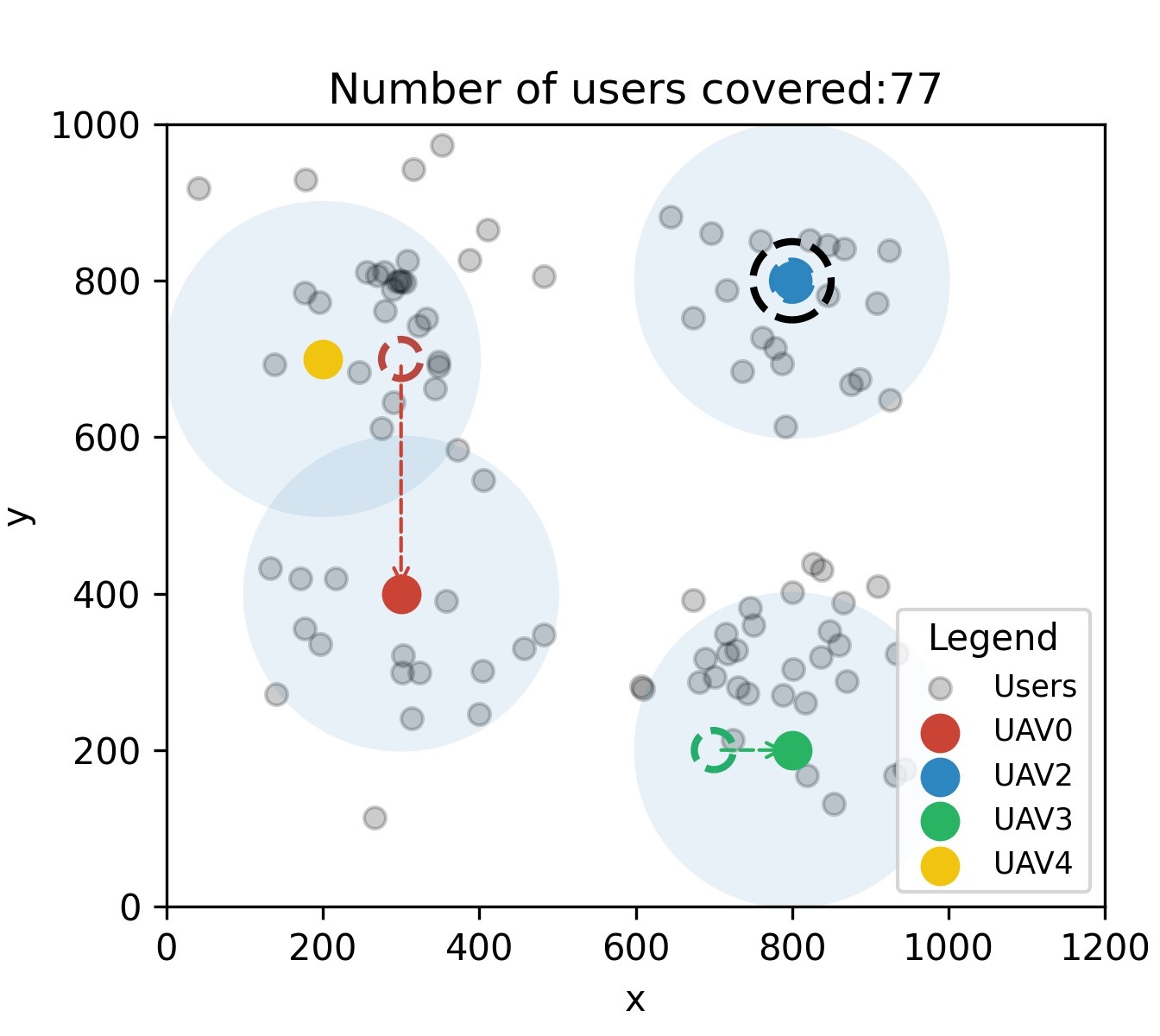}\label{Fig.fly3-22}}
\subfigure[1 UAV joins in] {\includegraphics[width=.45\linewidth]{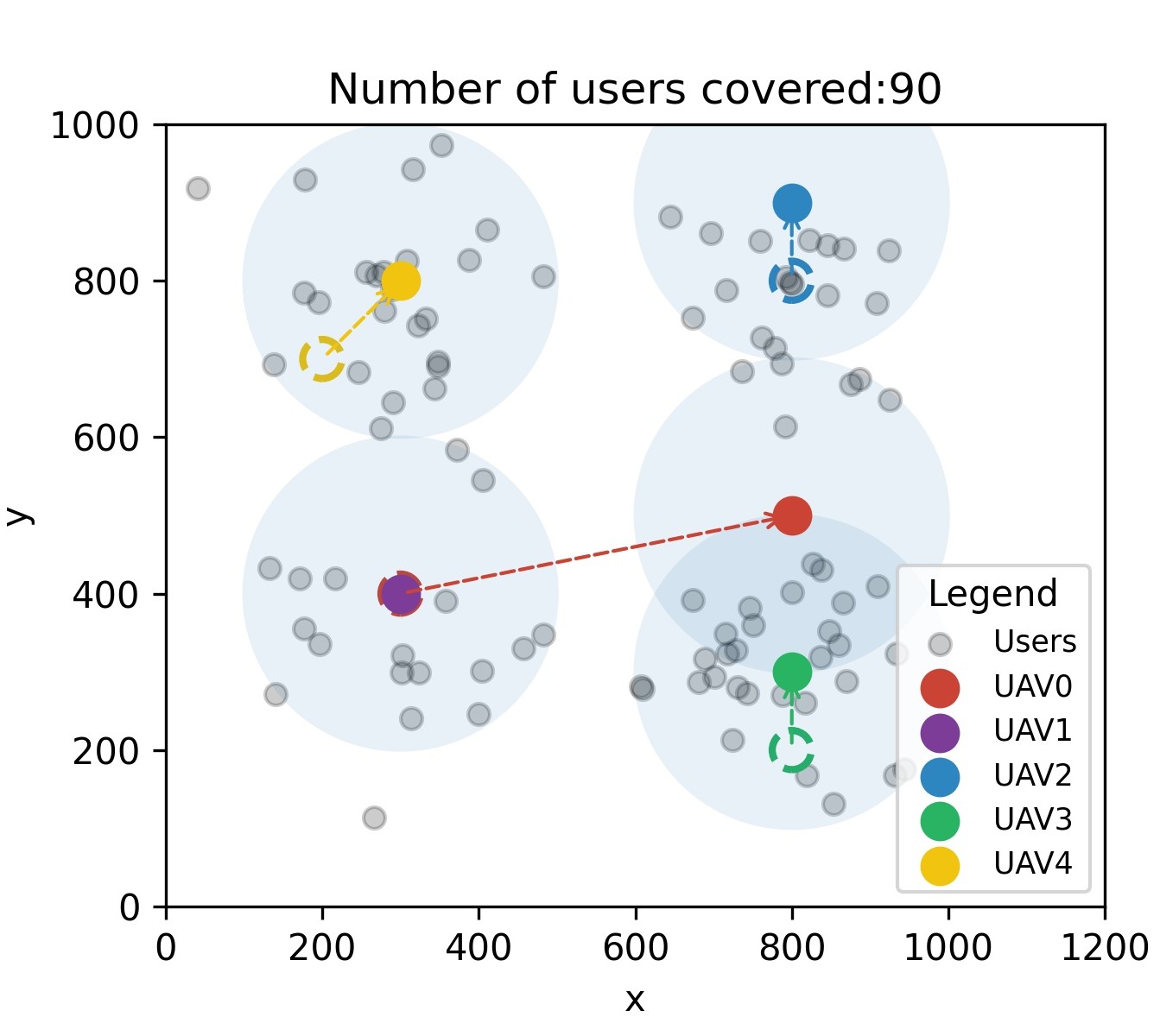}\label{Fig.fly3-33}}
\subfigure[1 UAV quits] {\includegraphics[width=.45\linewidth]{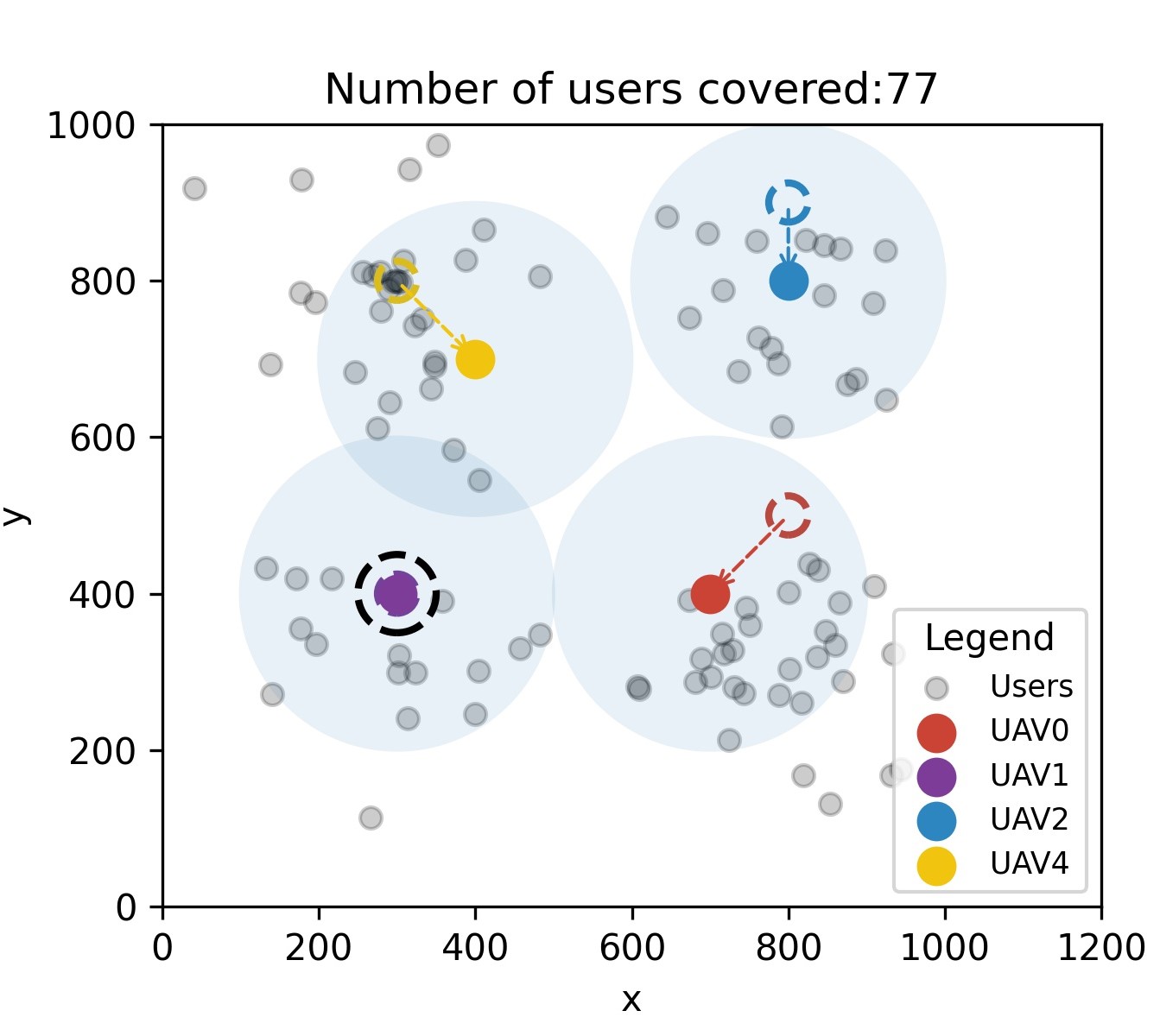}\label{Fig.fly3-44}}
\subfigure[1 UAV quits] {\includegraphics[width=.45\linewidth]{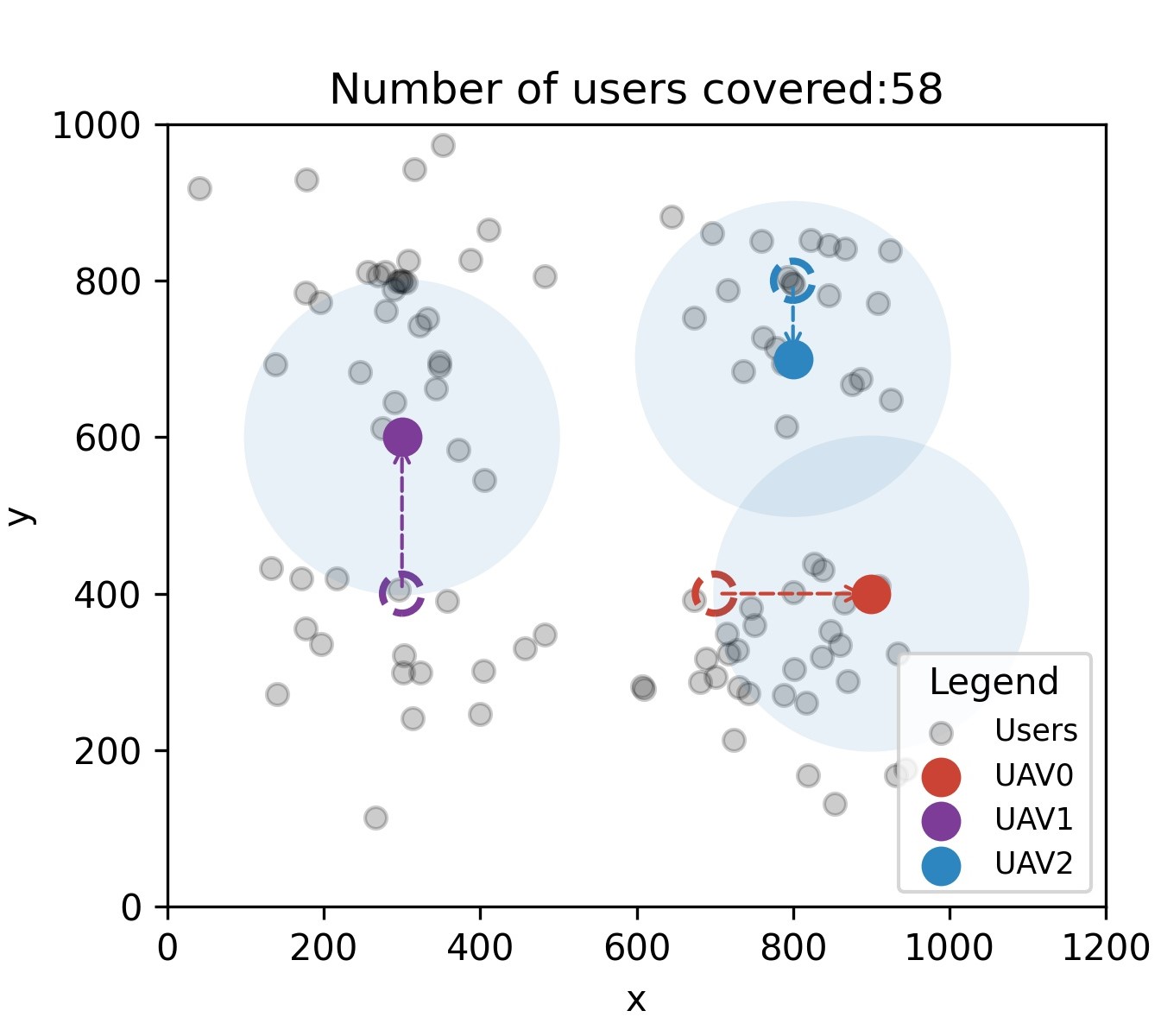}\label{Fig.fly3-55}}
\caption{Optimal coverage of active UAVs when UAVs randomly quit and join in sequentially}\label{fig:random1}
\end{figure}

The test results are given in Fig. \ref{Fig.testresults}. The results are grouped according to the actual number of connected users given the number of active UAVs. Each percentage is calculated according to the following example. For instance, in Fig. \ref{Fig.test2}, the percentage of 57 users being connected when there are 3 UAVs is 0.071. For all the 12 orders, we collect all the cases when 3 UAVs remain after a change (14 cases in total) and calculate the percentage of the cases with 57 users being connected (1 case in total). As a result, the corresponding percentage is 0.071. It can be seen that the obtained strategy is not absolutely optimal for every single order, yet is able to achieve less than 10$\%$ loss for over 90$\%$ of the cases. We believe a longer training time will improve the generality and possibly remove the outliners. For each type, one representative case is selected which is shown in Fig. \ref{fig:des2}-\ref{fig:random1}, respectively. In all the subfigures, the new positions of all the active UAVs after a quit/join-in are represented by solid dots of different colors, while their positions before the change are represented by hollow circles. It can be observed that every time when any UAV quits or joins in the network, the distributed policy is able to relocate the active UAVs to maximize the overall user connectivity. This is attributed to the strong adapatability of the obtained strategy under the deliberately designed structure of heterogeneous episodes.

\begin{figure}[!ht]
\centering
\subfigure[Random positions 1] {\includegraphics[width=.45\linewidth]{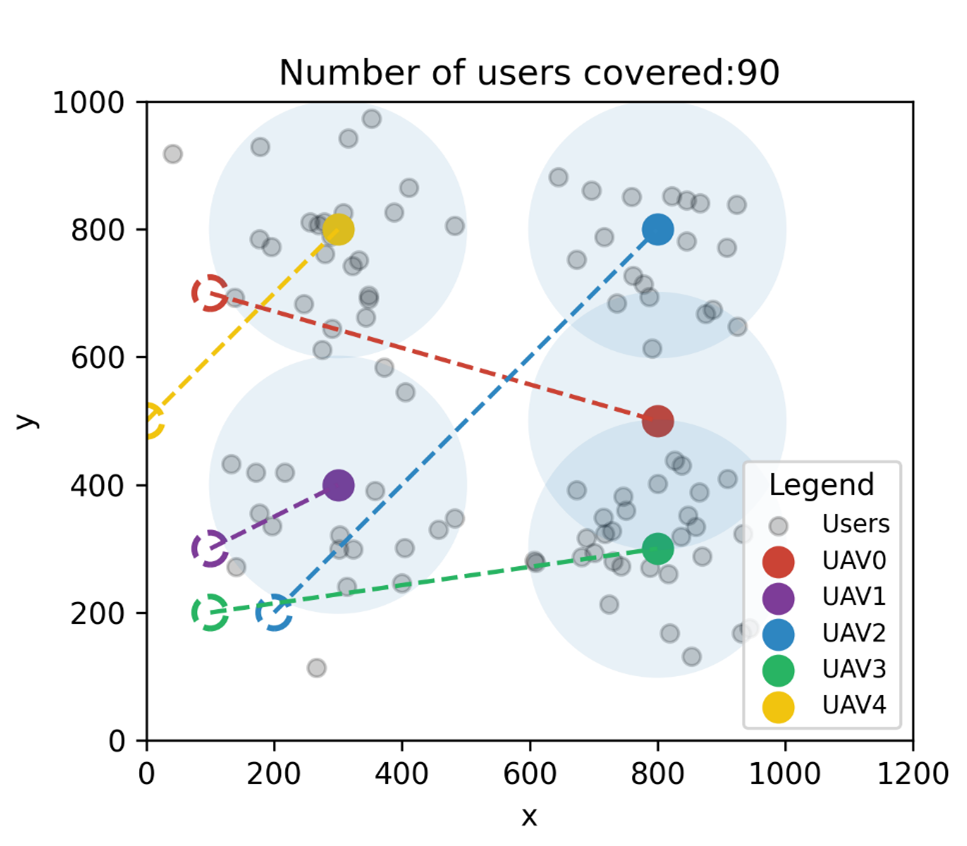}\label{Fig.random1}} 
\subfigure[Random positions 2] {\includegraphics[width=.45\linewidth]{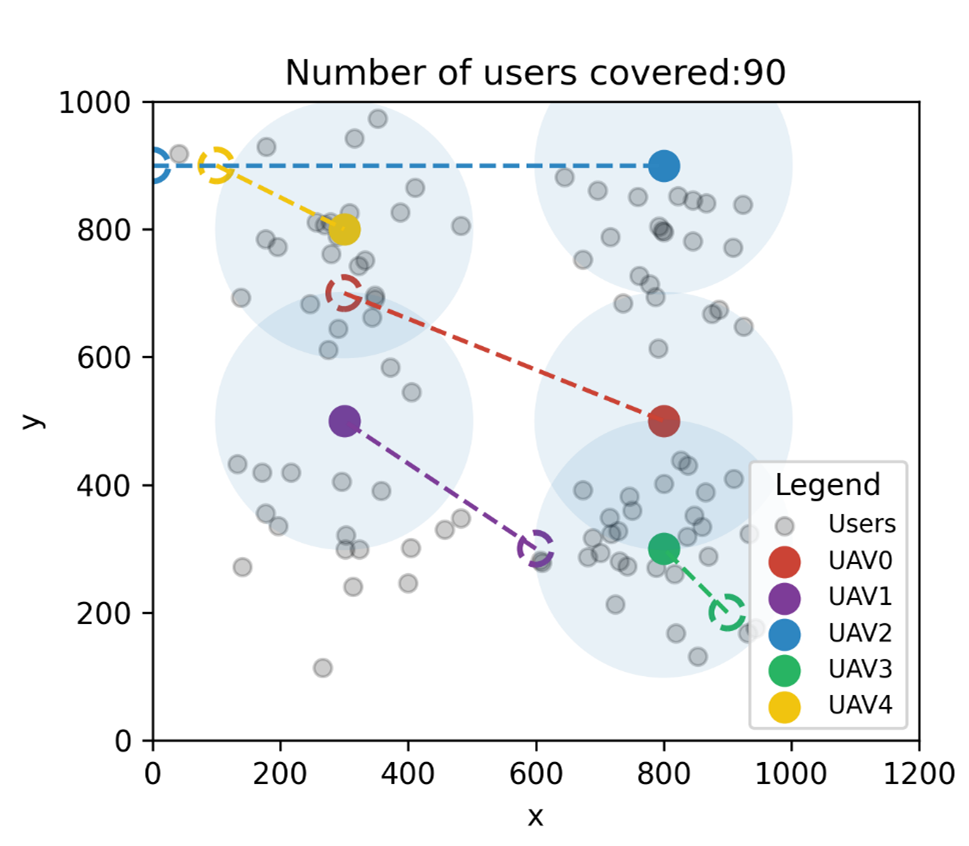}\label{Fig.random2}}
\subfigure[Random positions 3] {\includegraphics[width=.45\linewidth]{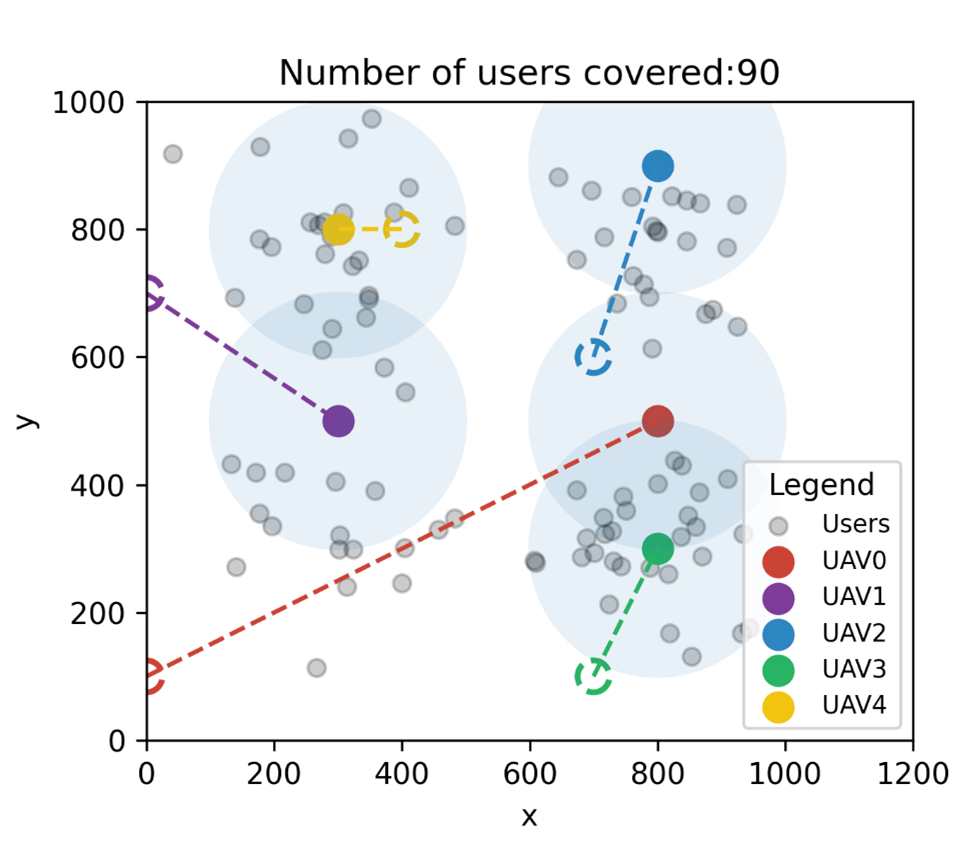}\label{Fig.random3}}
\caption{Illustrations demonstrating strong robustness against random initial UAV positions.}\label{fig:random}
\end{figure}

We then test the robustness of the achieved distributed policy against random initial UAV positions. We expect the policy to guide the UAVs to the optimal positions no matter where they start from. Fig. \ref{fig:random} shows 3 different sets of initial positions for 5 UAVs. It can be seen that for all the three sets, the achieved policy is able to guide all the UAVs to their respective optimal positions that yield the maximal overall user connectivity, i.e., 90. These serve as a strong proof that the achieved policy has strong robustness against random initial UAV positions, even though a fixed set of initial positions are used throughout the entire training.  

\section{Conclusions}\label{sec.Conclusion}
In this paper, a distributed user connectivity maximization problem has been studied in a UCN. The problem has been formulated into a time-coupled mixed integer non-convex optimization problem. Two algorithms, i.e., DUCM-1 and DUCM-2 algorithms, have been developed under the MARL framework to tackle the original problem in a distributed way. The developed algorithms have designed 4 different levels of info-sharing and evaluated their impact on the learning convergence. The evaluation is expected to shed some light on tuning the tradeoff between convergence performance and implementation complexity. The designed algorithms have also well handled the dynamics of UAV set and user distribution, thus being more practically applicable to realistic scenarios. Extensive simulations have demonstrated the impact of info-sharing and the efficacy and robustness of the proposed algorithms against the above dynamics. 



\bibliographystyle{IEEEtran}
\bibliography{main}

\end{document}